\documentclass[12pt]{article}
\pdfoutput=1
\usepackage{amssymb,amsmath,slashed,latexsym}
\usepackage{xcolor}
\usepackage{tikz}
\usetikzlibrary{calc}
\usetikzlibrary{matrix}
\usepackage{bbold}
\usepackage[vcentermath]{youngtab}
\usepackage{cite}

\usepackage[T1]{fontenc}

\usepackage{color}
\definecolor{dark-gray}{gray}{0.20}
\definecolor{gray}{gray}{0.30}
\definecolor{light-gray}{gray}{0.80}
\definecolor{dark-red}{rgb}{0.7,0,0}
\definecolor{dark-green}{rgb}{0.1,0.4,0}
\definecolor{dark-blue}{rgb}{0.3,0.3,0.7}
\definecolor{light-blue}{rgb}{0.8,0.8,1}

\usepackage{hyperref}
\hypersetup{
	colorlinks=true,
	linkcolor=dark-blue,
	citecolor=dark-red,
	urlcolor=dark-green,
	linktoc=page
}

\def\trace{\mathop{\rm Tr}\nolimits}
\def\rme{{\rm e}}
\def\rmi{{\rm i}}

\newcommand{\hc}{{\rm h.c.}}
\newsavebox{\uuunit}
\sbox{\uuunit}
{\setlength{\unitlength}{0.825em}
	\begin{picture}(0.6,0.7)
	\thinlines
	\put(0,0){\line(1,0){0.5}}
	\put(0.15,0){\line(0,1){0.7}}
	\put(0.35,0){\line(0,1){0.8}}
	\multiput(0.3,0.8)(-0.04,-0.02){12}{\rule{0.5pt}{0.5pt}}
	\end {picture}}

\csname @addtoreset\endcsname{equation}{section}

\newcommand{\SU}{\mathop{\rm SU}}
\newcommand{\SO}{\mathop{\rm SO}}
\newcommand{\SL}{\mathop{\rm SL}}
\newcommand{\U}{\mathop{\rm {}U}}
\newcommand{\USp}{\mathop{\rm {}USp}}

\newcommand{\E}{\mathop{\rm {}E}}

\newcommand{\N}{\mathcal{N}}

\usepackage[percent]{overpic}
\usepackage{caption}
\usepackage{subcaption}

\newcommand{\dd}{{\rm d}}
\newcommand{\e}{{\rm e}}
 
\begin{document}  

\begin{titlepage}
 
\bigskip
\bigskip
\bigskip
\begin{center} 
\textbf{\Large  Precision Holography for $\mathcal{N}=2^{*}$ on $S^4$ \\[0.2cm] from type IIB Supergravity }

\bigskip
\bigskip
\bigskip
\bigskip
\bigskip
\bigskip

\textbf{Nikolay Bobev, Fri\dh rik Freyr Gautason, and Jesse van Muiden }
\bigskip

Instituut voor Theoretische Fysica, KU Leuven,\\ 
Celestijnenlaan 200D, B-3001 Leuven, Belgium
\vskip 5mm

\bigskip
\bigskip
\bigskip
\tt{nikolay.bobev,~ffg,~jesse.vanmuiden~~~@kuleuven.be}  \\
\end{center}

\bigskip
\bigskip
\bigskip
\bigskip

\begin{abstract}

\noindent  
\end{abstract}

We find a new supersymmetric solution of type IIB supergravity which is holographically dual to the planar limit of the four-dimensional $\mathcal{N}=2^*$ supersymmetric Yang-Mills theory on $S^4$. We study a probe fundamental string in this background which is dual to a supersymmetric Wilson loop in the $\mathcal{N}=2^*$ theory. Using holography we calculate the expectation value of this line operator to leading order in the 't Hooft coupling. The result is a non-trivial function of the mass parameter of the $\mathcal{N}=2^*$ theory that precisely matches the result from supersymmetric localization. 

\noindent 

\vfill

\end{titlepage}

\newpage

\setcounter{tocdepth}{2}
\tableofcontents

 \section{Introduction}
 \label{Sec: Introduction}
 
Supersymmetric localization is a valuable tool which offers many insights into the strongly coupled dynamics of some QFTs by rendering their path integral calculationally tractable. Exact calculations of physical observables in QFT are clearly of great importance and, among other things, have lead to numerous explicit confirmations of various dualities, see \cite{Pestun:2016zxk} for a recent review and a list of references. The plethora of exact results in supersymmetric localization offers the exciting possibility to extend our understanding of supergravity and string theory by exploiting the gauge/gravity duality. Our goal here is to study a concrete setup in which localization results in QFT can make successful contact with supergravity.
 
The QFT of interest can be thought of as a mass deformation of the four-dimensional ${\SU(N)}$ $\mathcal{N}=4$ SYM theory which preserves $\mathcal{N}=2$ supersymmetry. This theory is well-studied and is referred to as $\mathcal{N}=2^*$ SYM. When the theory is placed on the round four-sphere, $S^4$, Pestun showed that some physical observables can be computed successfully by employing supersymmetric localization \cite{Pestun:2007rz}. The localization procedure reduces the path integral of the $\mathcal{N}=2^*$ SYM theory to a finite-dimensional matrix integral. Although this is a great simplification, the explicit evaluation of this integral is a daunting task due to the contribution of instantons to the integrand. In the large $N$ limit this matrix model simplifies significantly and is amenable to a more explicit analysis, see \cite{Russo:2012ay,Buchel:2013id,Russo:2013qaa,Russo:2013kea,Chen:2014vka,Russo:2017ngf} as well as \cite{Russo:2013sba,Zarembo:2016bbk} for a review and a more complete list of references. Two observables one can study explicitly in the planar limit are the partition function (or free energy) of the theory on $S^4$ as well as vacuum expectation values (vevs) of supersymmetric Wilson lines in various representations of the gauge group. It should be emphasized that these observables are still highly non-trivial functions of the 't Hooft coupling, $\lambda$, which are currently only accessible numerically. In fact, as shown in \cite{Russo:2012ay,Russo:2013qaa,Russo:2013kea,Chen:2014vka}, the $\mathcal{N}=2^*$ theory exhibits an infinite number of quantum phase transitions as one varies $\lambda$ from weak to strong coupling. In the strong coupling regime it is possible to obtain analytic results for the free energy and Wilson line expectation values as a function of the dimensionless mass parameter, $ma$, of $\mathcal{N}=2^*$ on a four-sphere of radius $a$. It is natural to try to reproduce these results from the supergravity approximation of type IIB string theory by employing the gauge/gravity duality.

Indeed this has been successfully done in \cite{Bobev:2013cja} for the free energy of the $\mathcal{N}=2^*$ theory on $S^4$. A solution of $\mathcal{N}=8$ five-dimensional $\SO(6)$ gauged supergravity was found in \cite{Bobev:2013cja} and it was argued that this is the five-dimensional gravitational dual of $\mathcal{N}=2^*$ on $S^4$. In addition, the regularized on-shell action of this solution was found to agree with the free energy of the theory as a function of $ma$. This five-dimensional gravitational solution is however not suitable for calculating vevs of Wilson lines via holography. For that purpose one has to study probe strings and branes in a ten-dimensional solution of type IIB supergravity. A particularly simple example of an analytically computable observable is offered by the supersymmetric Wilson line in the fundamental representation of the gauge group discussed in \cite{Pestun:2007rz}. In the planar limit of the gauge theory and for $\lambda\gg1$ it was shown in \cite{Buchel:2013id} that the vev of this line operator is 
\begin{equation}\label{eq:Wvevintro}
 \ln W\left(\mathcal{C}\right) = \sqrt{\lambda \left( 1 + m^2a^2 \right)}\,.
\end{equation}
Here $\mathcal{C}$ is a closed contour along the great circle (or equator) of $S^4$.

Our goal in this paper is two-fold, first we would like to understand how the supergravity solution of \cite{Bobev:2013cja} uplifts to a solution of type IIB supergravity. This turns out to be a somewhat non-trivial task and necessitates the application of some recent results in the literature on consistent truncations, exceptional field theory, and generalized geometry. We solve the problem by employing the uplift formulae derived in \cite{Baguet:2015sma}. The new type IIB supergravity solution we construct can be thought of as a generalization of the Pilch-Warner solution \cite{Pilch:2000ue}, which is the holographic dual of the $\SU(N)$ $\mathcal{N}=2^*$ SYM theory on $\mathbb{R}^4$.

With this explicit solution at hand, our second goal is to reproduce the result for the Wilson loop expectation value in \eqref{eq:Wvevintro} by a bulk calculation. To this end we study probe fundamental strings in the new type IIB supergravity solution. We employ the Nambu-Goto action and find the minimal energy configuration for a string which has the circular Wilson loop profile, $\mathcal{C}$, on the $S^4$ boundary of the asymptotically AdS$_5$ solution. Via the holographic dictionary the appropriately regularized on-shell action of this string should be dual to the Wilson loop expectation value. Indeed, performing this calculation explicitly we arrive at the field theory localization result in  \eqref{eq:Wvevintro}. Our results constitute a non-trivial precision test of the gauge/gravity duality for a non-conformal gauge theory to leading order in $\sqrt{\lambda}$. 

It should be noted that the authors of  \cite{Buchel:2013id} managed to reproduce the result in \eqref{eq:Wvevintro} in the limit $ma\gg1$ using the type IIB supergravity solution of \cite{Pilch:2000ue}. This is possible since for $ma\gg1$ the radius of $S^4$ becomes large and using the flat-sliced domain wall solution of \cite{Pilch:2000ue} is a justified approximation. In the limit $ma \to 0$ the result in \eqref{eq:Wvevintro} reduces to the well-known expectation value of a circular supersymmetric Wilson loop in the conformal $\mathcal{N}=4$ SYM, see for example \cite{Berenstein:1998ij,Drukker:1999zq}. Our holographic result is valid for general values of the parameter $ma$ and thus provides a generalization of the analysis in   \cite{Buchel:2013id,Berenstein:1998ij,Drukker:1999zq}.

The mass parameter, $m$, in the $\mathcal{N}=2^*$ theory on $S^4$ is in general complex and this leads to some subtleties in the interpretation and analysis of the supergravity solutions we study. In particular we find that for general complex values of the parameter $ma$ the ten-dimensional supergravity saddle point we derive is complex. This is a general feature of holographic applications of supergravity for QFTs defined on a curved Euclidean space and we discus it in more detail in the main text of the paper.
 
In the next section we summarize some basic facts about the $\mathcal{N}=2^*$ theory on $S^4$ as well as the results from localization for the vev of supersymmetric Wilson lines in the planar limit of the theory. In Section~\ref{Sec: Five dimensional supergravity} we describe the five-dimensional supergravity solution dual to $\mathcal{N}=2^*$ on $S^4$. The uplift of this background to a new solution of type IIB supergravity is presented in Section~\ref{Sec: From five to ten dimensions}. We study probe fundamental strings in this ten-dimensional solution in Section~\ref{Sec: Wilson loops and strings} and show that the regularized on-shell action of the string is equal to the expectation value of the supersymmetric circular Wilson line in the $\mathcal{N}=2^*$ theory to leading order in $\sqrt{\lambda}$. We present our conclusions and outline several interesting avenues for future work in Section~\ref{sec:conclusions}. The paper contains four appendices where we collect some of the technical results needed for our supergravity calculations.

 \section{Field theory}
 \label{Sec: Field theory}

A convenient way to think of the $\mathcal{N}=2^*$ theory of interest here is as a supersymmetric mass deformation of the maximally supersymmetric four-dimensional $\N=4$ SYM theory. The field content is a gauge field\footnote{The gauge group is $\SU(N)$ and we suppress all gauge indices.} $A_{\mu}$, six real scalars $X_i$, and four fermions $\lambda_m$, which are all in the adjoint representation of the gauge group. The scalars and fermions are in the $\mathbf{6}$ and $\mathbf{4}$ representations of the $\SO(6)$ R-symmetry group, respectively. It is useful to organize this field content in $\N=2$ multiplets. We have  a vector multiplet  
\begin{align}\label{eq:vecmultdef}
 A_{\mu}\,, \quad \psi_1= \lambda_4\,, \quad \psi_2=\lambda_3\,, \quad Z_3 =\frac{1}{\sqrt{2}}\left( X_3 + \rmi X_6 \right)\,,
\end{align}
and a hypermultiplet
\begin{align}\label{eq:hypmultdef}
\lambda_i\,, \quad Z_i = \frac{1}{\sqrt{2}} \left( X_i + \rmi X_{i+3} \right)\,, \quad \text{where} \quad i=1,2\,.
\end{align}
Here we are interested in studying the $\mathcal{N}=2^*$ SYM theory in Euclidean signature on the round four-sphere, $S^4$, of radius $a$. The corresponding Lagrangian was derived in \cite{Pestun:2007rz} and reads\footnote{We follow the notation and conventions of \cite{Bobev:2013cja}.}
 \begin{align}\label{Eq: N=2* Lagrangian on S4}
 \mathcal{L}_{\N=2^*}^{S^4} = \mathcal{L}_{\N=4}^{S^4} + \mathcal{L}_{\N=4}^{a} + \mathcal{L}_{m}^{\mathbb{R}^4} + \mathcal{L}_{m}^{a}\,.
 \end{align}
The first term on the right hand side of \eqref{Eq: N=2* Lagrangian on S4} is the standard Lagrangian of $\N=4$ SYM in flat space but with all regular partial derivatives changed into covariant derivatives on $S^4$. We note also that since we are in Euclidean signature the complex conjugated fields should be treated as independent variables. We emphasize this by denoting the conjugate of $Z_i$ by $\tilde{Z}_i$, however we still use $|Z_i|^2$ to denote the combination $Z_i\tilde{Z}_i$. The second term in \eqref{Eq: N=2* Lagrangian on S4} denotes the conformal coupling of the scalars in the theory to the curvature of $S^4$
 \begin{align}\label{Eq: N=4 radius correction for S4}
 \mathcal{L}_{\N=4}^{a} = \frac{2}{a^2} \trace \left( |Z_1|^2 + |Z_2|^2 + |Z_3|^2 \right)\,.
 \end{align}
The third term in \eqref{Eq: N=2* Lagrangian on S4} is the standard mass deformation that breaks the $\N=4$ supersymmetry to $\N=2$. This term is the same as in flat space and reads:\footnote{The mass parameter $m$ is in general a complex number.}
 \begin{align}\label{eq:LmasR4}
 \mathcal{L}_{m}^{\mathbb{R}^4} = m^2 \trace \left(|Z_1|^2 + |Z_2|^2\right) + m \trace\left( \lambda_1^2 + \lambda_2^2 + \hc  \right)+m\left((\tilde{Z}_1Z_2-\tilde{Z}_2Z_1)Z_3 +\hc \right)\,.
 \end{align}
The last term in \eqref{Eq: N=2* Lagrangian on S4} was introduced in \cite{Pestun:2007rz} and is necessary to preserve $\mathcal{N}=2$ supersymmetry on $S^4$ 
 \begin{align}\label{eq:LmasS4}
 \mathcal{L}_{m}^{a} = \frac{\rmi m}{2 a} \trace\left( Z_1^2 + Z_2^2 + \hc \right)\,.
 \end{align}
In this $\mathcal{N}=2$ notation only an ${\rm SU(2)}_V\times {\SU(2)}_H\times {\rm U(1)}_R$ subgroup of the ${\SO(6)}$ R-symmetry group of the $\mathcal{N}=4$ SYM theory is manifest. The mass deformation in \eqref{eq:LmasR4} completely breaks ${\rm U(1)}_R$ and preserves only a ${\U(1)}_H$ subgroup of ${\SU(2)}_H$. The deformation in \eqref{eq:LmasS4} preserves only a ${\rm U(1)}_V$ subgroup of ${\rm SU(2)}_V$. Therefore we conclude that the Lagrangian of the $\mathcal{N}=2^*$ theory on $S^4$ preserves ${\rm U(1)}_V\times {\U(1)}_H$ global symmetry. 

It was noted in \cite{Intriligator:1998ig} that $\mathcal{N}=4$ SYM has an extra continuous symmetry in the planar limit, namely the compact ${\rm U(1)}_s$ subgroup of the ${\rm SL}(2,\mathbb{R})$ S-duality group. It was then observed in \cite{Pilch:2000ue,Buchel:2000cn} that the diagonal ${\rm U(1)}_Y$ subgroup of ${\rm U(1)}_s$ and ${\rm U(1)}_R$ is also preserved in the planar limit of the $\mathcal{N}=2^*$ theory on $\mathbb{R}^4$. The evidence for this comes from the holographic dual description of the theory and was shown to hold also for the $\mathcal{N}=2^*$ theory on $S^4$ in \cite{Bobev:2013cja}. Our supergravity results below provide further evidence for the invariance of the $\mathcal{N}=2^*$ theory under this bonus ${\rm U(1)}_Y$ symmetry.

 \subsection{Results from localization}
 \label{subsec:Locresults}
  
One can use supersymmetric localization to show that the path integral for the $\mathcal{N}=2^*$ SYM theory on $S^4$ reduces to a matrix integral over the vevs of the real scalar field $X_3$ \cite{Pestun:2007rz}. This integral is in general hard to evaluate explicitly due to the presence of non-trivial instanton contributions captured by Nekrasov's partition function. The calculation is under better control in the large $N$ limit when one can argue that instantons do not contribute and the matrix integral becomes manageable \cite{Russo:2012ay}. Despite these drastic simplifications it is still not known how to explicitly evaluate the path integral of the  $\mathcal{N}=2^*$ SYM theory on $S^4$ as a function of 't Hooft coupling $\lambda$ and $ma$ in the limit $N\gg 1$. For general values of $\lambda$ one should resort to numerics and as shown in a series of papers by Russo and Zarembo the $\mathcal{N}=2^*$ SYM theory on $S^4$ in the planar limit possesses a rich phase structure, see \cite{Russo:2012ay,Russo:2013qaa,Russo:2013kea,Chen:2014vka,Russo:2013sba}. In the limit $\lambda \gg 1$ one can solve the problem analytically and find that the free energy of the theory is
\begin{equation}\label{eq:Fintro}
  F_{S^4} = -\frac{N^2}{2} (1 + m^2 a^2) \log \frac{\lambda (1 + m^2 a^2) e^{2 \gamma + \frac 12}}{16 \pi^2} \,.
\end{equation}
As discussed in \cite{Buchel:2013id,Bobev:2013cja,Bobev:2016nua} this quantity is scheme dependent. One way to obtain a scheme independent observable is to differentiate \eqref{eq:Fintro} three times with respect to the dimensionless variable $ma$, see \cite{Bobev:2013cja,Bobev:2016nua} for a more detailed discussion. 

Another interesting set of physical observables computable by supersymmetric localization is given by Wilson loop operators in various representations of the gauge group. Of main interest in this paper is the expectation value of the following supersymmetric Wilson line operator in the fundamental representation of the gauge group
\begin{align}\label{eq:WLdef}
W(\mathcal{C}) = \left\langle \frac{1}{N} \mathcal{P} {\rm exp}\left[\oint_{\mathcal{C}} \dd t \left( \rmi A_{\mu} \partial_t x^{\mu} + |\partial_t x| X_3 \right)\right] \right\rangle\,.
\end{align}
Here $t$ parametrizes the contour $\mathcal{C}$ given by the great circle on $S^4$ with stereographic coordinates $x^{\mu}$. The real scalar $X_3$ is defined in \eqref{eq:vecmultdef} and is singled out by the supersymmetric localization calculation as the only scalar in $\mathcal{N}=2^*$ which has a non-zero expectation value \cite{Pestun:2007rz}. As shown in \cite{Buchel:2013id} in the planar limit of $\mathcal{N}=2^*$, and in the strong coupling regime $\lambda\gg 1$, the expectation value of the Wilson loop in \eqref{eq:WLdef} is given by the simple expression
\begin{equation}\label{eq:Wvev}
 \ln W\left(\mathcal{C}\right) = \sqrt{\lambda \left( 1 + m^2a^2 \right)}\,.
\end{equation}
One of our goals in this paper is to reproduce this result from the type IIB supergravity limit of string theory. This can be achieved by studying probe fundamental strings in a non-trivial classical solution of the supergravity theory that is the holographic dual of the $\mathcal{N}=2^*$ theory on $S^4$. We now proceed to describe how to construct this supergravity solution.

 \section{Five-dimensional supergravity}
 \label{Sec: Five dimensional supergravity}

Since the $\mathcal{N}=2^*$ theory is a deformation of four-dimensional $\mathcal{N}=4$ SYM it is natural to expect that its holographic dual description can be constructed as a deformation of the familiar AdS$_5\times S^5$ solution of type IIB supergravity. Finding this solution directly in ten dimensions is a difficult task since the $\SO(6)$ isometry of $S^5$ is broken to $\U(1)_V\times \U(1)_H$. A more fruitful approach is to use the fact that the lowest lying KK modes on $S^5$ are captured by the five-dimensional $\mathcal{N}=8$ ${\rm SO(6)}$ gauged supergravity theory. The solution of interest is then realized in five dimensions as a deformation of the maximally supersymmetric AdS$_5$ vacuum by non-trivial profiles for some of the scalar fields in the theory. Using this approach the gravity dual description of the $\mathcal{N}=2^*$ SYM theory on $\mathbb{R}^4$ and $S^4$ was found in \cite{Pilch:2000ue} and \cite{Bobev:2013cja}, respectively. Below we summarize the construction of the solution in \cite{Bobev:2013cja}.

The maximal ${\rm SO(6)}$ gauged five-dimensional supergravity was constructed in \cite{Gunaydin:1984qu,Gunaydin:1985cu,Pernici:1985ju}. The bosonic fields in the theory are the metric, 42 scalar fields, 12 2-forms, and the ${\rm SO(6)}$ gauge field. Since we are interested in solutions which are deformations of AdS$_5$ and preserve the Poincar\'e (for $\mathcal{N}=2^*$ on $\mathbb{R}^4$) or ${\rm SO(5)}$ (for $\mathcal{N}=2^*$ on $S^4$) invariance of the boundary, only the metric and the scalar fields in the theory can be non-trivial. 
 
The five-dimensional gauged supergravity is invariant under the maximal subgroup, $\SL(6,\mathbb{R})\times \SL(2,\mathbb{R})$, of the  $\E_{6(6)}$ symmetry enjoyed by the ungauged supergravity and all the fields in the theory are in representations of this maximal subgroup. The 42 scalars of interest to us parametrize the coset $\E_{6(6)}/\USp(8)$. As explained in \cite{Gunaydin:1985cu} one can parametrize this coset space, and thus the $42$ scalars of the supergravity theory, by the $27\times 27$ matrix
 \begin{align}\label{Eq: Xhat matrix}
 \hat{X} = \left( \begin{array}{cc}
 -4 \Lambda_{[I}^{\phantom{[I}[P}\delta_{J]}^{Q]}  & \sqrt{2} \Sigma_{IJR\beta} \\ 
 \sqrt{2}\Sigma^{PQK\alpha} & \Lambda_R^{\phantom{[I}K}\delta_{\beta}^{\alpha} + \Lambda_{\beta}^{\phantom{\beta}\alpha} \delta^{K}_{R}
 \end{array}  \right).
 \end{align}
Here the capital Latin indices transform under $\SL(6,\mathbb{R})$ and the lower-case Greek indices transform under $\SL(2,\mathbb{R})$. The matrix $\Lambda_{I}^{\phantom{I}J}$ is $6\times 6$ symmetric and traceless, $\Lambda_{\alpha}^{\phantom{\alpha}\beta}$ is $2\times 2$ symmetric and traceless, and the tensor $\Sigma_{IJK\alpha}$ is completely antisymmetric and self-dual in the indices $IJK$ for $\alpha=1,2$ and thus has $20$ independent components.

To obtain the action and supersymmetry variations of the supergravity theory one has to work with a group element of the scalar coset manifold given by the tensor $U = \rme^{\hat{X}}$. This $U$ tensor can be thought of as the vielbein on the scalar manifold. For our purposes it turns out to be more convenient to work with the metric on the coset space given by
 \begin{align}\label{eq:Mdef}
 M = U\cdot U^T~,
 \end{align}
where $U$ is transposed and multiplied directly as a $27\times 27$ matrix.  The elements of the matrix $M$ can be further split up into representations of $\SL(6,\mathbb{R})\times \SL(2,\mathbb{R})$ as follows
\begin{align}\label{Eq: Scalar matrix}
M=
\left(\begin{matrix}
M_{IJ,PQ} & M_{IJ}^{\phantom{IJ}R\beta} \\ 
M^{K \alpha}_{\phantom{K\alpha}PQ} & M^{K \alpha, R \beta}
\end{matrix}\right)\,.
\end{align}
Here the capital Latin index pairs $IJ$ and $PQ$ should be treated as antisymmetric and therefore transform in the ${\bf 15}$ of $\SL(6,\mathbb{R})$. The index pair involving one Latin index and one Greek are in the $({\bf 6},{\bf 2})$ of $\SL(6,\mathbb{R})\times \SL(2,\mathbb{R})$ and so can take twelve different values. In practice the decomposition in the right hand side of \eqref{Eq: Scalar matrix} is therefore a simple block decomposition of the $27\times27$ matrix on the left hand side into $15\times15$, $15\times12$ and $12\times12$ blocks. Knowing explicitly the components of the matrix $M$ is an essential ingredient in applying the uplift formulae of \cite{Baguet:2015sma}.

The scalar and gravity sector of the five-dimensional supergravity theory is described by the following Lagrangian in Euclidean signature
 \begin{align}\label{Eq: Baguet:2015sma Lagrangian}
 \mathcal{L} = \left(R_5 + \frac{1}{24} \text{tr}(\partial_{\mu} M \cdot \partial^{\mu} M^{-1}) - V \right) ,
 \end{align}
where the trace and matrix multiplication is over the ${\bf 27}$ indices of $M$. To calculate the scalar potential one also needs to use as input the matrix $\hat{X}$ in \eqref{Eq: Xhat matrix}. We summarize how this calculation proceeds in Appendix~\ref{Sec: Some properties of the exceptional group E66}, see also \cite{Gunaydin:1985cu}.

 \subsection{The three-scalar model}
 \label{subsec:3scalar}

To construct the five-dimensional supergravity dual of the $\mathcal{N}=2^*$ SYM theory on $S^4$ one can take advantage of the global symmetries of the theory described below \eqref{eq:LmasS4}. Imposing the $\U (1)_V\times\U (1)_H\times \U (1)_Y$ symmetry on the scalars of the supergravity theory results in a consistent truncation with only four independent scalars, see \cite{Bobev:2013cja}. In addition one can consistently set one of these scalars to zero. The three remaining scalars, which we call $\alpha$, $\beta$, and $\chi$, are precisely the supergravity duals of the three relevant operators which trigger the RG flow from $\mathcal{N}=4$ SYM to the $\mathcal{N}=2^*$ theory 
 \begin{align}\label{eq:Oabchidef}
 \mathcal{O}_{\alpha} = \trace\left(|Z_1|^2 + |Z_2|^2\right)\,, \quad \mathcal{O}_{\beta} = \trace \left( Z_1^2 + Z_2^2 + \hc \right)\,, \quad
 \mathcal{O}_{\chi} =  \trace\left( \lambda_1^2 + \lambda_2^2 + \hc  \right).
 \end{align}

The scalars $\alpha$ and $\beta$ belong to the $\mathbf{20'}$ representation of $\SO(6)$ and are specified by the following matrix $\Lambda_{J}^{\phantom{J}I}$ in \eqref{Eq: Xhat matrix}
 \begin{equation}\label{eq:Lambda3def}
 \Lambda_{J}^{\phantom{J}I}=\left(\begin{array}{cccccc}
 -\alpha + \beta &  &  &  &  &  \\ 
 & -\alpha - \beta &  &  &  &  \\ 
 &  & -\alpha + \beta &  &  &  \\ 
 &  &  & -\alpha - \beta &  &  \\ 
 &  &  &  & 2\alpha &  \\ 
 &  &  &  &  & 2\alpha
 \end{array}\right) .
 \end{equation}
The scalar $\chi$ sits in the $\mathbf{10}\oplus \mathbf{\overline{10}}$ representation of $\SO(6)$ and the non-zero elements of the tensor $\Sigma_{IJK\alpha}$ in \eqref{Eq: Xhat matrix} are given by
 \begin{equation}\label{eq:Sigma3def}
 \Sigma_{IJ 5 2} = -\Sigma_{IJ 6 1} = \left(\begin{matrix}
 &  & \chi &  &  &  \\ 
 &  &  & -\chi &  &  \\ 
 -\chi &  &  &  &  &  \\ 
 & \chi &  &  &  &  \\ 
 &  &  &  &  &  \\ 
 &  &  &  &  & 
 \end{matrix} \right).
 \end{equation}
 There are no other scalars in the truncation and thus the matrix $\Lambda_{\beta}^{\phantom{\beta}\alpha}$ vanishes.
 
To write down the bosonic Lagrangian of this truncation it is convenient to introduce the following notation
\begin{equation}
\eta = \e^{\alpha}\,, \qquad z = \frac{\tanh\sqrt{\beta^2+\chi^2}}{\sqrt{\beta^2+\chi^2}}(\beta+\rmi\chi)\,, \qquad \tilde z = \frac{\tanh\sqrt{\beta^2+\chi^2}}{\sqrt{\beta^2+\chi^2}}(\beta-\rmi\chi)\,.
\end{equation}
We should note here that since we are interested in supergravity solutions in Euclidean signature the complex scalar $z$ should be treated as independent from its formal complex conjugate $\tilde{z}$. We adopt the notation $\tilde{z}$ to emphasize this fact. Plugging the tensors in \eqref{eq:Lambda3def} and \eqref{eq:Sigma3def} into the $\hat{X}$ matrix in \eqref{Eq: Xhat matrix} one finds that the Lagrangian in \eqref{Eq: Baguet:2015sma Lagrangian} reads
 \begin{equation}\label{Eq: Bobev:2013cja Lagrangian}
 {\cal L}= \left[ R_5 - \frac{12 \partial_{\mu}\eta\partial^{\mu}\eta}{\eta^2} - 4\frac{\partial_{\mu}z \partial^{\mu}\tilde z}{(1-z\tilde z)^2}- V\right]\,,
 \end{equation}
where
\begin{equation}\label{Eq: Bobev:2013cja potential}
 V = -g^2 \left(\frac{1}{\eta^4} + 2 \eta^2 \frac{1+ z\tilde z}{1-z\tilde z} - \frac{\eta^8}{4} \frac{(z-\tilde z)^2}{\left(1-z\tilde z^2\right)^2}\right)\,.
 \end{equation}
The constant $g=2/L_{{\rm AdS}}$ in \eqref{eq:5dmetric} is the gauge coupling of the five-dimensional supergravity and sets the scale, $L_{{\rm AdS}}$, of the maximally supersymmetric AdS$_5$ solution in the theory. 

The five-dimensional solution of interest is a domain wall with $S^4$ slices captured by the following metric Ansatz
\begin{equation}\label{eq:5dmetric}
\dd s^2_5 = \dd r^2 + e^{2A(r)}\dd \Omega_{4}^2\,.
\end{equation}
Here $\dd \Omega_{4}^2$ is the metric on the round $S^4$ given explicitly in \eqref{eq:S4met}. The three scalars are functions only of the radial coordinate $r$. One can plug this Ansatz in the supersymmetry variations of the maximal supergravity theory and derive a set of BPS equations which determine the solutions of interest. This was described in \cite{Bobev:2013cja} to which we refer for more details. The BPS equations take the form of three non-linear ODEs for the scalar fields in the model given by 
\begin{equation}\label{Eq: BPS equations for C and T}
\begin{aligned}
z' =& -3(1-z\tilde z)(\eta')\frac{2(z+\tilde z)+\eta^6(z-\tilde z)}{2\eta(1+\tilde z^2 -(1-\tilde z^2)\eta^6)}~,\\
\tilde z' =& -3(1-z\tilde z)(\eta')\frac{2(z+\tilde z)-\eta^6(z-\tilde z)}{2\eta(1+ z^2 -(1- z^2)\eta^6)}~,\\
(\eta')^2 =& g^2\frac{(1+ z^2 -(1- z^2)\eta^6)(1+\tilde z^2 -(1-\tilde z^2)\eta^6)}{36(1-z\tilde z)^2\eta^2}\,.
\end{aligned}
\end{equation}
In addition, there is an algebraic constraint which determines the metric function $A(r)$ in terms of the scalars
\begin{align}\label{eq:e2Aalg}
\e^{A} = \frac{12 (1-z\tilde z)^2(\eta')}{g^2 (z^2-\tilde z^2)\eta^3}\,.
\end{align}
This algebraic equation is consistent with the following differential equation for $A$
\begin{equation}\label{eq:Adiff}
 A'=	g^2\frac{2 \left(1+z^2\right)\left(1+\tilde{z}^2\right)-\eta^6 \left(1-z^2 \tilde{z}^2\right)-\eta ^{12} \left(1-z^2\right) \left(1-\tilde{z}^2\right)}{36 \eta ^3 (\eta ') (1-z \tilde{z})^2}~.
\end{equation}
This equation for $A'(r)$ proves to be useful when we take a limit of the three scalar model appropriate for gravitational domain walls with flat slices in the metric, i.e. with a metric Ansatz as in \eqref{eq:5dmetric} but with the metric on $\mathbb{R}^4$ instead of $S^4$. In this limit, discussed in more detail in Appendix~\ref{app:PWSTU}, the algebraic equation in \eqref{eq:e2Aalg} does not hold and we must resort to the differential one. It is important to note that the BPS equations above are invariant under two independent discrete symmetries, namely  the exchange  $z \leftrightarrow \tilde{z}$ as well as $\{z,\tilde{z}\} \to -\{z,\tilde{z}\}$. Solutions related by these symmetries are therefore physically equivalent.

\subsection{Solving the BPS equations}
\label{subsec:5dsol}

Unfortunately we are not able to solve the differential equations in \eqref{Eq: BPS equations for C and T} analytically. Following \cite{Bobev:2013cja} we resort to a linearized analysis in the UV and IR and a numerical solution of the system in \eqref{Eq: BPS equations for C and T}.

The metric in \eqref{eq:5dmetric} is asymptotic to $\mathbb{H}^5$ (or Euclidean AdS$_5$) and the boundary is at large values of the coordinate $r$ in \eqref{eq:5dmetric}. This corresponds to the UV regime of the gauge theory. One can linearize the BPS equations in this limit and find the approximate solution
\begin{equation}\label{eq:UV5d}
\begin{aligned}
\eta &\approx 1 + \left(\frac{\mu^2}{3}g r+\frac{\mu(\mu+v)}{3}\right) \e^{-gr}\,,\\
\frac{1}{2}(z+\tilde{z}) &\approx  \left(\mu g r+v\right) \e^{-gr} \,,\\
\frac{1}{2}(z-\tilde{z}) &\approx \mp \mu \e^{-g r/2} \mp \left(\frac{2}{3}\mu(\mu^2-3)gr + \frac{1}{3}\left(2v(\mu^2-3)+\mu(4\mu^2-3)\right)\right) \e^{-3gr/2}\,,\\
\e^{2A} &\approx \frac{\e^{gr}}{g^2} +\frac{2}{3g^2}(\mu^2-3) \,.
\end{aligned}
\end{equation}
The higher order term in this linearized solution are fully determined by the two integration constants $\mu$ and $v$. This UV expansion matches the expectation from the gauge theory. As discussed in Section~\ref{Sec: Field theory} all three operators in $\eqref{eq:Oabchidef}$ are added to the Lagrangian of $\mathcal{N}=4$ SYM and their coefficients are related to each other due to supersymmetry. This is manifested in the BPS equations since the leading source terms in \eqref{eq:UV5d} are controlled by the parameter $\mu$. This parameter is related to the dimensionless deformation parameter $ma$ in the gauge theory and it was shown in \cite{Bobev:2013cja} that the relation is 
\begin{equation}\label{eq:muimadef}
\mu = \pm \rmi ma\;.
\end{equation}
The freedom to choose a sign in \eqref{eq:UV5d} and \eqref{eq:muimadef} is due to the discrete symmetries of the BPS equations discussed at the end of Section~\ref{subsec:3scalar}.

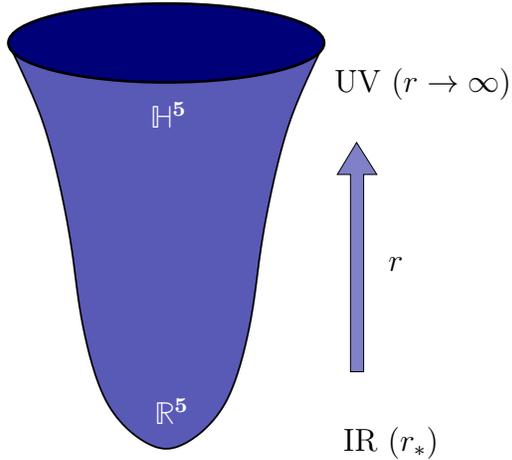
\begin{figure}[h]
\centering
\begin{overpic}[width=0.3\textwidth,tics=10]{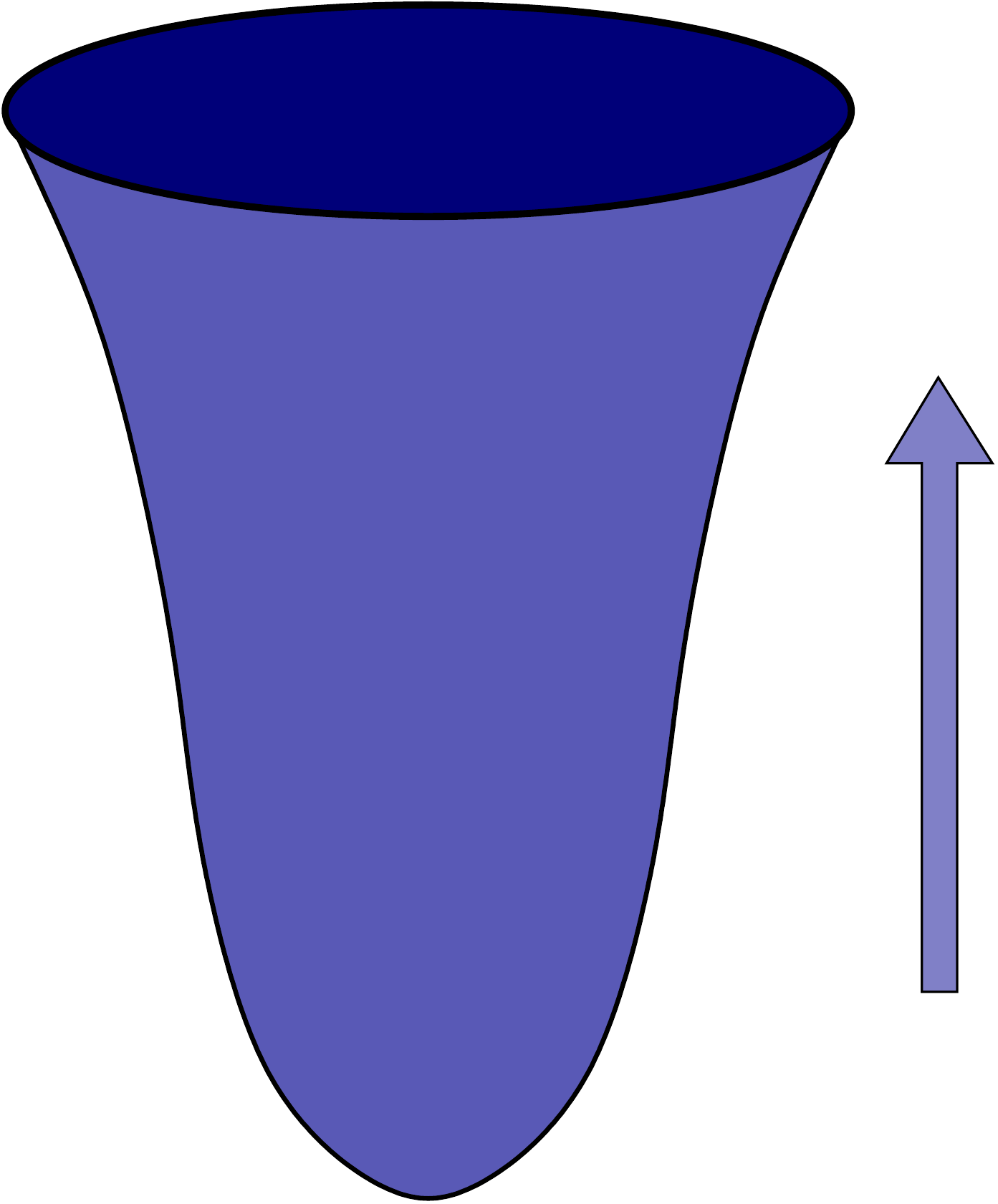}
\put (85,40) {$r$}
\put (75,0) {IR ($r_*$)}
\put (73,80) {UV $(r\to \infty)$}
\put (32,72) {\textcolor{white}{\textbf{$\mathbb{H}^\mathbf{5}$}}}
\put (33,6) {\textcolor{white}{$\mathbf{\mathbb{R}^5}$}}
\end{overpic}
\caption{\label{cap}The regular five-dimensional geometries interpolate between AdS$_5$ in the UV and $\mathbb{R}^5$ in the IR.}
\end{figure}
The radius of $S^4$ provides an IR cutoff for the dynamics of the gauge theory. This in turn selects a unique vacuum on the moduli space of the $\mathcal{N}=2^*$ theory, see \cite{Russo:2013sba} for a nice discussion of the vacuum selection mechanism. It is natural to expect that in the supergravity solution dual to $\mathcal{N}=2^*$ on $S^4$ this vacuum selection mechanism is manifested by a regularity condition in the core of the geometry. It was indeed found in \cite{Bobev:2013cja} that the BPS equations in \eqref{Eq: BPS equations for C and T} and \eqref{eq:e2Aalg} admit such a regular solution. This is achieved by selecting some value $r=r_{*}$ where the $S^4$ shrinks smoothly to zero size and the metric is simply $\mathbb{R}^5$, see Figure \ref{cap}.\footnote{The particular value of $r_*$ is not a physical parameter since we can change it by shifts of the coordinate $r$. When writing down the UV expansion \eqref{eq:UV5d}, we have made use of this shift symmetry to eliminate a similar non-physical parameter.} This regularity constraint leaves only one integration constant and the BPS equations have the approximate solution
\begin{equation}\label{eq:IR5d}
\begin{aligned}
\eta &\approx \eta_0 -\frac{\eta_0^{12}-1}{108\eta_0^3} g^2 (r-r_*)^2\,,\\
\frac{1}{2}(z+\tilde{z}) &\approx \sqrt{\frac{\eta_0^6-1}{\eta_0^6+1}} \left(\frac{\eta_0^6}{\eta_0^6+2} - \frac{ \eta_0^8(4\eta_0^6+5)}{30(\eta_0^6 +2)^2}\, g^2(r-r_*)^2  \right)\,,\\
\frac{1}{2}(z-\tilde{z}) &\approx \mp \sqrt{\frac{\eta_0^6-1}{\eta_0^6+1}} \left(\frac{2}{\eta_0^6+2}   +\frac{ \eta_0^2 (3 \eta_0^{12}- 10 \eta_0^6 -20)}{60(\eta_0^6 +2)^2}\, g^2(r-r_*)^2 \right) \,,\\
\e^{2A} &\approx (r-r_{*})^2 \,.
\end{aligned}
\end{equation}
The higher order terms in this linearized IR solution are determined algebraically by the single integration constant $\eta_0$. The sign freedom in the third equation in \eqref{eq:IR5d} is again due to the discrete symmetries in the BPS equations. For $\eta_0=1$ the solution is simply $\mathbb{H}^5$ and the scalars do not flow.

\begin{figure}[h]
\centering
\includegraphics[width=0.6\textwidth,tics=10]{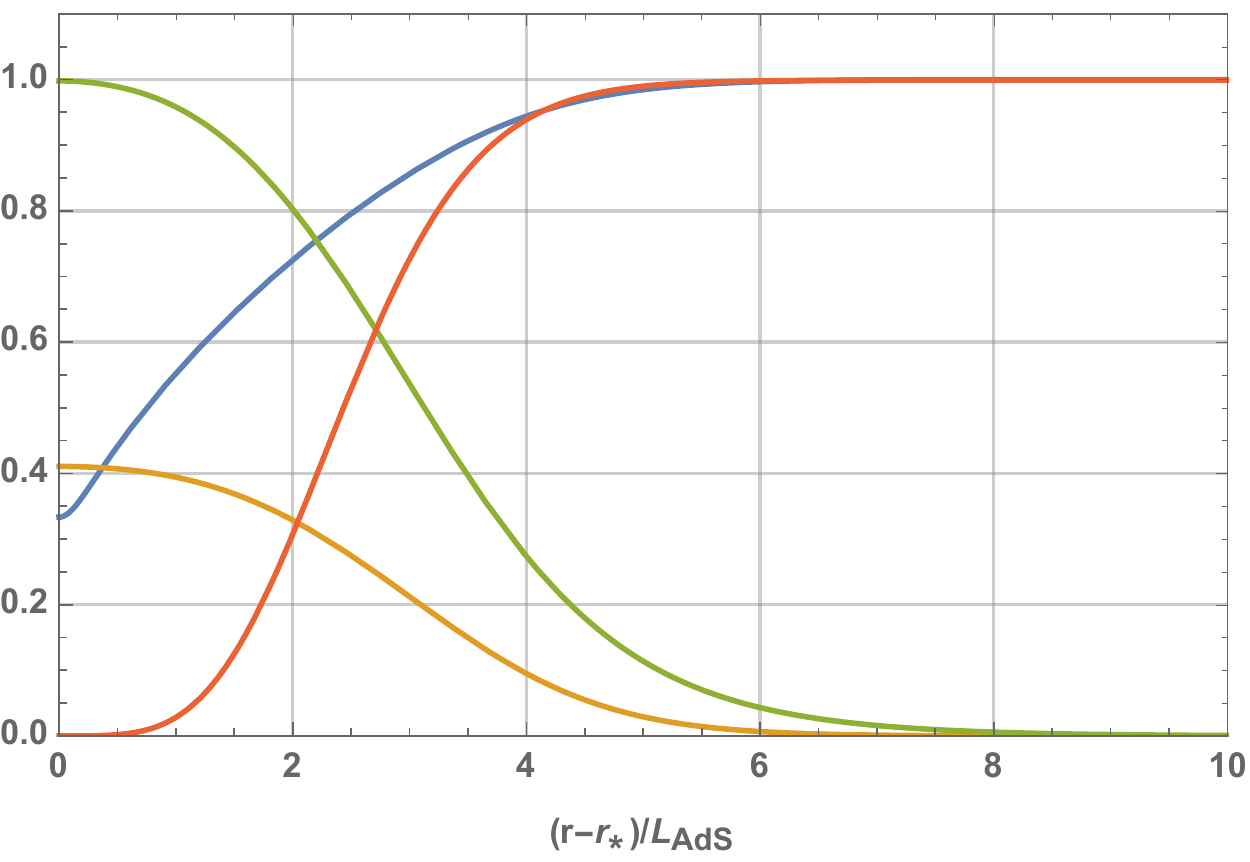}
\caption{\label{shooting}An example of a numerical solution to the BPS equations (\ref{Eq: BPS equations for C and T}--\ref{eq:Adiff}). The solution was obtained by shooting from the IR with $\eta_0 = 1/3$ and other fields as in \eqref{eq:IR5d}. The blue curve is $\eta$, the yellow one is $300(z+\tilde z)$, the green one $(\tilde z-z)/2$, and the red one is $g~\e^{2A-gr}/2$.}
\end{figure}
The full nonlinear solution of the BPS equations can be obtained by numerically integrating the BPS equations and interpolating between the IR and UV expansions above. We implement this numerical procedure by ``shooting'' from the IR at $r_*$ and integrating the nonlinear ODEs up to large values of $r$ where the solution can be matched onto the expansion in \eqref{eq:UV5d}. This is performed for numerous different values of the IR parameter $\eta_0$ which leads to a one-parameter family of numerical BPS solutions. A representative solution is depicted in Figure~\ref{shooting}. In general the parameter $\eta_0$ is complex which then results in complex values for the metric function $A(r)$. For the detailed numerical analysis in this paper we have chosen to focus on real values of $\eta_0\neq 0$ since this leads to real metric functions $A(r)$.\footnote{We have numerically integrated the BPS equations for hundreds of different complex values of $\eta_0$ and found regular solutions with complex metric functions.} For real $\eta_0$ there are two branches of solutions to the system of BPS equations, one for $|\eta_0|>1$ and one for $|\eta_0|<1$. Since $\eta$ is real, the equations in \eqref{Eq: BPS equations for C and T} imply that $\eta$ is monotonic throughout the flow. This means that in order to reach the UV where $\eta\to 1$ we must select the correct branch of the square root of $(\eta')^2$. Similarly the BPS equations imply that when $|\eta|>1$ both $z$ and $\tilde z$ are real whereas for $|\eta|<1$, $z$ and $\tilde z$ are pure imaginary. Correspondingly, the UV integration constants $\mu$ and $v$ in \eqref{eq:UV5d} are either both real or both imaginary depending on the sign of $|\eta|-1$. We should also stress that due to the regularity condition in the IR we have only one integration constant, $\eta_0$, and thus we should find a relation between the two integration constants $\mu$ and $v$ which control the UV expansion. Indeed such a relation was found numerically in \cite{Bobev:2013cja} where it was shown that the numerical results are in very good agreement with the analytic relation\footnote{We have checked numerically that this relation holds not only for real or pure imaginary values of $\mu$ but more generally on the complex plane.}
\begin{equation}\label{eq:vmurelation}
v(\mu) = -2\mu-\mu\log(1-\mu^2)\,.
\end{equation}
We will use these numerical results when we discuss the uplift of this five-dimensional solution to ten-dimensional type IIB supergravity.

\section{From five to ten dimensions}
\label{Sec: From five to ten dimensions}

It has long been suspected that the five-dimensional $\SO (6)$ $\mathcal{N}=8$ gauged supergravity  is a consistent truncation of type IIB supergravity on $S^5$. This was established rigorously only recently in \cite{Lee:2014mla,Baguet:2015sma}. The utility of this consistent truncation is that one can take any solution of the five-dimensional supergravity theory and uplift it to a solution of the ten-dimensional type IIB supergravity. To this end the explicit uplift formulae in \cite{Baguet:2015sma} (see also \cite{Hohm:2013vpa,Baguet:2015xha}) are very useful. We will not describe in any detail how these formulae are obtained and will simply apply them to uplift the five-dimensional solution described in the previous section.
 
The only information from the five-dimensional theory that one needs in order to uplift the solution of \cite{Bobev:2013cja} to ten dimensions is the metric in \eqref{eq:5dmetric} and the scalar matrix in \eqref{Eq: Scalar matrix}, as well as some basic facts about the geometry of $S^5$ that are summarized in Appendix~\ref{Sec: Coordinates}.

\paragraph{The dilaton/axion} The dilaton, $\Phi$, and axion, $C^{(0)}$, of type IIB supergravity parametrize an $\SU(1,1)/\U(1)$ coset space. The $\SU(1,1)$ matrix, $m^{\alpha \beta}$, is given by the following formula
\begin{align}\label{Eq: Uplift formula dilaton/axion}
m^{\alpha \beta} = \Delta^{4/3} Y_I Y_J M^{I \alpha, J \beta}\,.
\end{align}
The function $\Delta$ is determined by imposing unimodularity, i.e. $\det(m^{\alpha \beta})=1$. The embedding coordinates, $Y_I$, of $S^5$ in $\mathbb{R}^6$ are given explicitly in Appendix~\ref{Sec: Coordinates}. The axion and dilaton can be extracted from the matrix $m^{\alpha\beta}$ by using
\begin{align}
m^{\alpha\beta} = \left(
\begin{array}{cc}
e^{\Phi}
(C^{(0)})^2+e^{-\Phi} & -e^{\Phi} C^{(0)} \\
-e^{\Phi} C^{(0)} & e^{\Phi} \\
\end{array}
\right)\,.
\end{align}

\paragraph{The metric} The internal space is topologically $S^5$ but due to the non-trivial values of the five-dimensional scalar fields, the metric on $S^5$ is not the Einstein one and is determined by the formula
\begin{align}\label{eq:Guplift}
G^{mn} =  \mathcal{K}_{IJ}^{\phantom{IJ}m} \mathcal{K}_{PQ}^{\phantom{PQ}n} M^{IJ,PQ}\,,
\end{align}
where $\mathcal{K}_{IJ}^{\quad m}$ are the Killing vectors on the round five sphere, see Appendix~\ref{Sec: Coordinates} for an explicit expression of these vectors in terms of the embedding coordinates $Y_I$. Here the indices $m,n=1,\dots,5$ refer to the directions on the five sphere. It is important to notice that the components of the matrix $M$ that appear here and in \eqref{eq:A2uplift} below are not components of the matrix \eqref{Eq: Scalar matrix} but rather its inverse. In addition, the five-dimensional non-compact metric in \eqref{eq:5dmetric} acquires a warp factor which depends on the coordinates of $S^5$ and is determined by the function $\Delta$ in \eqref{Eq: Uplift formula dilaton/axion}. The full ten-dimensional metric in Einstein frame takes the form
\begin{equation}
\dd s_{10}^2 = \Delta^{-2/3}\left(\dd s_5^2 + \dd\Omega_5^2\right)\,,
\end{equation}
where $\dd\Omega_5^2$ is the line element for the metric on the deformed $S^5$ given by \eqref{eq:Guplift}.

\paragraph{The two-forms} To obtain the ten-dimensional NS-NS two-form $B^{(2)}$ and the R-R two-form $C^{(2)}$ it is convenient to define $A^{\phantom{mn}1}_{mn} = B_{mn}^{(2)}$ and $A^{\phantom{mn}2}_{mn} = C^{(2)}_{mn}$. One then has 
\begin{align}\label{eq:A2uplift}
A_{mn}^{\quad \alpha} = -\frac{1}{g}\varepsilon^{\alpha \beta}  {G}_{nk} ~\mathcal{K}_{IJ}^{\quad k} M^{IJ}_{\quad P \beta} ~\partial_{m} Y^P\,,
\end{align}
where ${G}_{mn}$ is the deformed metric of the five sphere given in \eqref{eq:Guplift}. The corresponding three-form field strengths are defined in Appendix~\ref{app:IIBsugra}. The expression on the right hand side of \eqref{eq:A2uplift} is not manifestly antisymmetric in the indices $m$ and $n$. However, using the properties of the matrix $M$ and the Killing vectors $\mathcal{K}$ one can show that it is indeed antisymmetric.

\paragraph{The four-form} The components of the R-R four-form along the $S^5$ are given by
\begin{align}\label{eq:C4uplift}
C_{klmn}^{(4)} = \frac{4}{g^4}\left( \sqrt{\widehat G} ~\varepsilon_{klmnp}  ~\widehat G^{pq} \Delta^{4/3}m_{\alpha \beta}\partial_{q}(\Delta^{-4/3}m^{\alpha \beta})+\widehat{\omega}_{klmn}\right)\,,
\end{align} 
where $\widehat{G}_{mn}$ is the metric on the round $S^5$ given in \eqref{Eq: Round five sphere metric} and $\widehat{G}$ is its determinant. The four-form $\widehat\omega$ is such that
\begin{equation} \label{eq:domegadef}
\dd \widehat\omega =16\,{\rm vol}_{S^5}= 8\sin2\theta\cos^2\theta \,\dd\theta\wedge \dd\phi \wedge \sigma_1\wedge \sigma_2\wedge \sigma_3\,.
\end{equation}
The right hand side of \eqref{eq:domegadef} is proportional to the volume form of the metric on the round $S^5$ in the coordinates defined in \eqref{Eq: Round five sphere metric}. The $\sigma_i$ are the left invariant one forms of $\SU(2)$, the explicit parametrization we chose for these forms is given in \eqref{eq:sig_i}. To obtain the five-form flux of type IIB supergravity from the four-form in \eqref{eq:C4uplift} one should use the formula in \eqref{Eq: RR fields on internal space}.
 
\subsection{A new solution of type IIB supergravity}
\label{Sec: A new solution of type IIB supergravity}

Applying these uplift formulae to the five-dimensional model described in Section \ref{Sec: Five dimensional supergravity} results in a new supersymmetric solution of type IIB supergravity. It is convenient for our purposes to present the solution by first defining the following two combinations of the scalar fields $z$ and $\tilde{z}$
\begin{align}\label{eq:CTdef}
C = \frac{1+z\tilde z}{1-z\tilde z}~,\qquad T = \frac{z+\tilde z}{1+z\tilde z}\,.
\end{align}
The metric on the internal $S^5$ is given by
\begin{align}\label{eq:S5deformed}
\dd {\Omega}_5^2 =& \frac{4}{g^2 \eta^2}\left(\frac{1}{CS}\dd\theta^2+ \frac{1}{K_2}\sin^2\theta~\dd\phi^2+ \eta^6\cos^2\theta\left[\frac{S}{K_1}(E_1^2+E_2^2) + \frac{1}{C S K_2 }E_3^2\right] \right)\,,
\end{align}
where we have used $\dd\Omega_5^2$ to denote the \emph{deformed} metric on $S^5$, see \eqref{eq:Guplift}. We refer to Appendix \ref{Sec: Coordinates} for a discussion on how we chose the coordinates on the deformed sphere. The one forms $E_i$ are deformations of the $\SU(2)$ left invariant one-forms $\sigma_i$ and are given by
 \begin{align}
 E_1 =& \sigma_1 +\frac{T \sin \omega}{S}\left(\tan\theta\sin\xi_{-}\dd\theta + \cos\xi_{-}\sigma_3\right)\,,\nonumber\\
 E_2 =& \sigma_2 +\frac{T \sin \omega}{S}\left(\tan\theta\cos\xi_{-}\dd\theta - \sin\xi_{-}\sigma_3\right)\,,\\
 E_3 =& \sigma_3\,. \nonumber
 \end{align}
We have also found it convenient to define
 \begin{align}\label{Eq: Scalar functions}
	S = &~ 1 +T \cos\omega\,, \notag\\
	K_1 =&~ S \cos^2\theta + C(1-T^2) \eta^6 \sin^2\theta\,,\\
	K_2 =&~ S C \cos^2\theta + \eta^6\sin^2\theta \,. \notag
\end{align}
Using these definitions, the full ten-dimensional metric in Einstein frame can be written compactly as
\begin{align}\label{eq:10dmetric}
\dd s^2_{10} = \frac{(C K_1 K_2)^{1/4}}{\eta\sqrt{g_s}}(\dd s_5^2 + \dd\Omega_5^2)\,, 
\end{align}
where $\dd s^2_5$ is the five-dimensional metric in \eqref{eq:5dmetric} and $\dd\Omega_5^2$ is given by \eqref{eq:S5deformed}. Notice that we have introduced the string coupling constant $g_s$ into the ten-dimensional solution by hand.  The dilaton and axion are given by
\begin{equation}\label{eq:10dPhiC0}
\begin{aligned}
\e^{\Phi} =& \frac{g_s}{\sqrt{C K_1 K_2}} (C K_1 \sin^2 \phi + K_2 \cos^2 \phi)\,,\\ 
C^{(0)} =& \frac{\eta^6\left(C^2 \left(1-T^2\right)-1\right)\sin ^2\theta  \sin 2\phi}{2g_s\left(C K_1 \sin^2\phi+K_2 \cos^2\phi\right)}\,.
\end{aligned}
\end{equation}
The NS-NS and R-R two-forms can be written as
\begin{equation}\label{eq:B2C2IIB}
B^{(2)} = -\cos\phi~\Psi_2 + \sin\phi~\dd\phi\wedge \Psi_1~,\qquad g_sC^{(2)} = \sin\phi~\Psi_2 + \cos\phi~\dd\phi\wedge\Psi_1~,
\end{equation}
where we have defined
\begin{equation}
\begin{aligned}
\Psi_2 =& \frac{2\sqrt{C^2 \left(1-T^2\right)-1}}{ g^2}\cos \theta \left[ \frac{2}{ C S}  \dd\theta\wedge E_3 - \frac{ S \eta^6}{K_1}\sin 2\theta \,E_1 \wedge E_2\right]~,\\ 
\Psi_1 =& \frac{2\sqrt{C^2 \left(1-T^2\right)-1}}{ g^2K_2}\cos \theta \sin 2\theta ~ E_3~.\\
\end{aligned}
\end{equation}
The four-form along the the internal $S^5$ is given by 
\begin{align}
C^{(4)} =&\frac{8S(C K_1 + K_2)}{g_s g^4K_1K_2}\cos^4\theta~E_1\wedge E_2\wedge E_3\wedge\dd\phi\,. 
\end{align}

We have checked explicitly that the bosonic fields above obey the equations of motion of type IIB supergravity, which are given in Appendix \ref{app:IIBsugra}. To perform this consistency check we have used the BPS equations in \eqref{Eq: BPS equations for C and T} and \eqref{eq:e2Aalg}. Notice however, that the equations of motion in Appendix \ref{app:IIBsugra} are presented in Lorentzian signature whereas our solution is Euclidean in ten dimensions. Despite this, we still find a consistency of the BPS equations with the ten-dimensional equations of motion because some of the ten-dimensional fields are pure imaginary (or more generally complex). For example, when $|\eta|<1$, the metric, the axion-dilaton as well as the four-form are all real. However, the two-forms $B^{(2)}$ and $C^{(2)}$ are pure imaginary. Redefining the two-forms by multiplying them by the imaginary unit must be accompanied by changing the ten-dimensional equations of motion to the ones of Euclidean type IIB supergravity. More generally when the ten-dimensional fields are complex we still find a consistency with the Lorentzian equations of motion even though a simple redefinition of fields does not result in a solution of the Euclidean theory. In Appendix~\ref{app:PWSTU} we show how, after taking appropriate limits of the five-dimensional scalar fields, this background reduces to two well-known analytic solutions of type IIB supergravity.

The supergravity background above can be thought of as sourced by D3-branes. This implies that the five-form flux through the deformed $S^5$ should be appropriately quantized. This leads to the following relation between the number of D3-branes, $N$, and the other parameters in the solution
\begin{equation}
N = \frac{4}{g^4 g_s \ell_s \pi}\,,
\end{equation}
here $\ell_s$ is the string length. Furthermore for large values of the coordinate $r$ the geometry above correctly reduces to AdS$_5\times S^5$ with length scales
\begin{equation}\label{eq:constants}
L_\text{AdS}^2 = L_{S^5}^2 =\frac{4}{g^2} = \sqrt{4N g_s \pi}~\ell_s^2 = \sqrt{\lambda}~\ell_s^2\,,
\end{equation}
where we have used that the 't Hooft coupling is $\lambda = g_\text{YM}^2 N$, and that the Yang-Mills and string couplings are related via $g_\text{YM}^2= 4\pi g_s$.

As discussed above \eqref{eq:vmurelation}, for general complex values of the IR parameter $\eta_0$ all three scalars and the metric function $A(r)$ are in general complex. However when $\eta_0$ is real the five-dimensional metric is also real. The behavior of the ten-dimensional solutions is somewhat more involved. For $|\eta_0|<1$ we find that both $C$ and $T$, as defined in \eqref{eq:CTdef}, are real, therefore the ten-dimensional metric is real. For $|\eta_0|>1$ we find that the scalar combination $T$ is imaginary and thus the ten-dimensional background fields are complex.

\subsection{Symmetries of the solution}

The metric \eqref{eq:10dmetric} exhibits $\SO(5)\times U(1)^3$ symmetry as expected from the field theory discussion in Section \ref{Sec: Field theory}. The $\SO(5)$ symmetry is generated by the isometries of the four-sphere present in the five-dimensional metric $\dd s_5^2$. The $\U(1)^3$ symmetries are generated by the Killing vectors
\begin{equation}
\partial_{\xi_+}\,,\qquad \partial_{\xi_-}\,,\qquad \partial_{\phi}\,.
\end{equation}
The first of these correspond to the $\U(1)_V$ symmetry of the dual field theory, the second corresponds to the $\U(1)_H$  whereas the last one is dual to $\U(1)_R$. However, as discussed below \eqref{eq:LmasS4} the $\U(1)_R$ is broken by the mass terms \eqref{eq:LmasR4} and \eqref{eq:LmasS4}. Indeed we find that the two-forms in \eqref{eq:B2C2IIB} as well as the axion and dilaton in \eqref{eq:10dPhiC0} break the $\U(1)$ symmetry generated by $\partial_{\phi}$. This symmetry is however restored when a shift of $\phi$ is combined with a particular element of the $\SL(2,\mathbb{R})$ symmetry group of type IIB supergravity. Under a general $\SL(2,\mathbb{R})$ transformation the R-R and NS-NS 2-forms and the axion-dilaton, $\tau = C^{(0)} + {\rm i} \e^{-\Phi}$, transform as
\begin{equation}
\left(\begin{matrix} C^{(2)}\\ B^{(2)} \end{matrix}\right) \mapsto \left(\begin{matrix} a&b\\c&d \end{matrix}\right)\left(\begin{matrix} C^{(2)}\\ B^{(2)} \end{matrix}\right)\,,\qquad \tau\mapsto \frac{a\tau + b}{c\tau +d}\,,\qquad ad-bc =1\,.
\end{equation}
We therefore see that a shift $\phi\mapsto \phi+\delta$ together with the $\U(1)\subset\SL(2,\mathbb{R})$ rotation generated by 
\begin{equation}
\left(\begin{matrix} \cos\delta&\sin\delta\\-\sin\delta&\cos\delta \end{matrix}\right)~,
\end{equation}
leaves the 2-forms  as well as the axion and dilaton invariant. This invariance is the supergravity manifestation of the bonus $\U(1)_Y$ symmetry discussed below \eqref{eq:LmasS4}.

 \section{Holographic Wilson loops}
 \label{Sec: Wilson loops and strings}
 
The explicit ten-dimensional solution constructed above can be used to study the dynamics of probe strings and branes which via the holographic dictionary can be mapped to expectation values of line and surface operators. Our main interest here is to study probe fundamental strings which are holographically dual to Wilson lines in the fundamental representation in the gauge theory \cite{Maldacena:1998im,Rey:1998ik}.
 
The expectation value of a Wilson line operator defined along a contour ${\cal C}$ can be calculated holographically by evaluating the renormalized on-shell action for a string with a world-sheet that ends on the contour ${\cal C}$ at the $\mathbb{H}^5$ boundary and extends in the bulk of the supergravity solution \cite{Maldacena:1998im,Rey:1998ik}. More precisely 
\begin{align}\label{Eq: General Wilson loop dictionary}
\log W(\mathcal{C}) = -S_\text{string}^R\,,
\end{align}
where $S_\text{string}^R$ is the renormalized on-shell action of the fundamental string given by the sum of the Nambu-Goto action and the coupling to the NS-NS $B^{(2)}$ field
 \begin{align}\label{eq:Sstrdef}
 S_\text{string} = \frac{1}{2\pi \ell_s^2}\int \dd^2 \sigma\; \e^{\Phi/2}\sqrt{\det P[G_{M N}]} - \frac{1}{2\pi \ell_s^2}\int P\left[B^{(2)}\right]\,.
 \end{align}
Here $P[\cdots]$ denotes the pull-back of the corresponding bulk fields onto the string world-sheet parametrized by coordinates $\sigma^1$ and $\sigma^2$ and we use the ten-dimensional metric $G_{MN}$ in Einstein frame. When calculating the on-shell value of the string action, we make use of the probe approximation and neglect any backreaction of the string on the background. The calculation then amounts to minimizing the string action, regularizing it, and evaluating it on-shell. 
 
\begin{figure}[h]
\centering
\begin{overpic}[width=0.3\textwidth,tics=10]{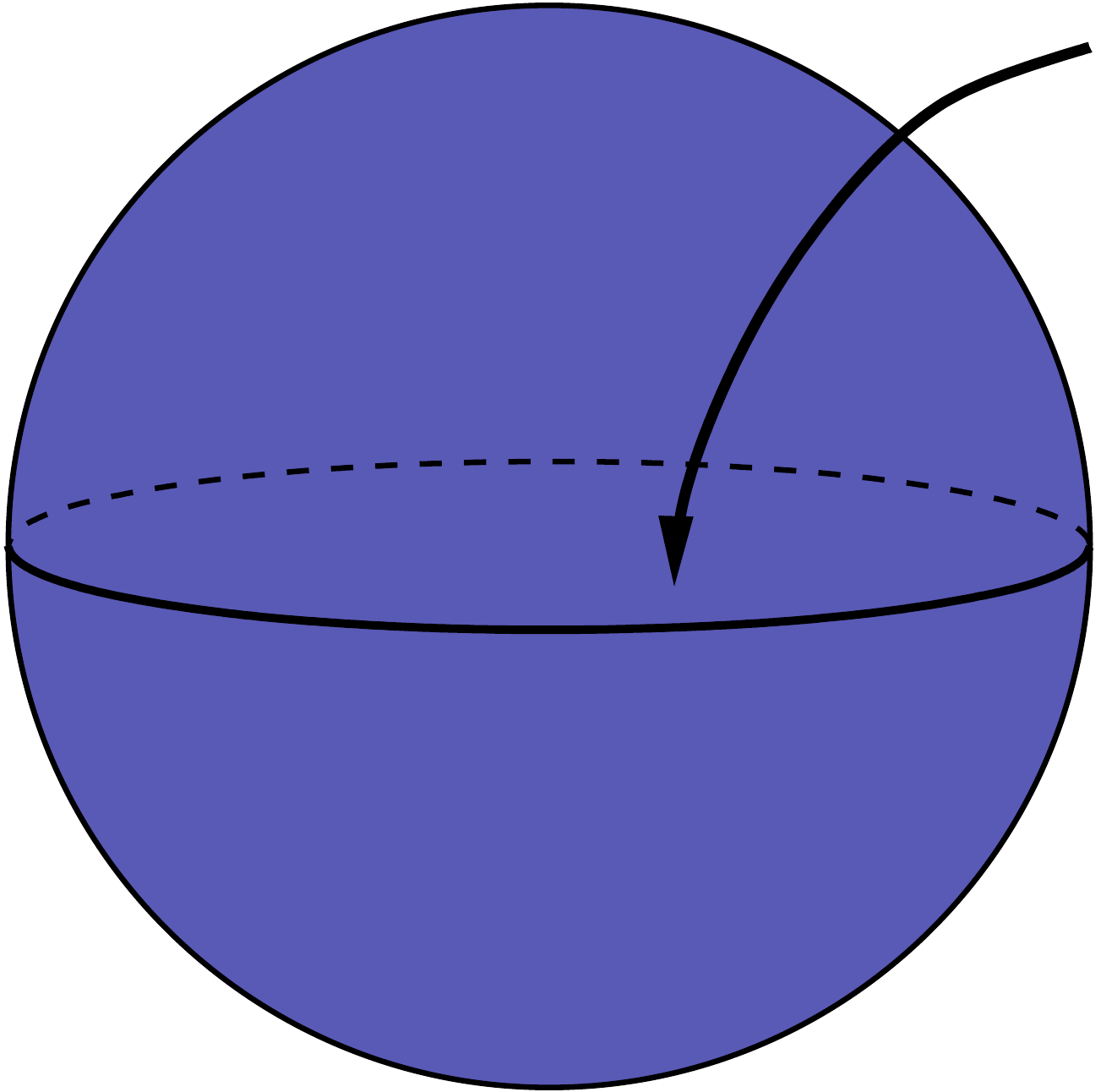}
\put (102,92) {\large$\mathcal{C}$}
\put (0,85) {\LARGE$S^4$}
\end{overpic}
\caption{\label{sphere}The contour $\mathcal{C}$ that defines the Wilson line is a great circle on $S^4$.}
\end{figure}
The Wilson loop operator we consider is the one defined in \eqref{eq:WLdef}. The contour $\mathcal{C}$ is along the great circle of $S^4$, see Figure~\ref{sphere}, which we parametrize with the coordinate $\sigma^1=t \in [0,2\pi]$.\footnote{The parameter $t$ can be identified with the coordinate $\zeta_4$ in \eqref{eq:S4met}.} The other coordinate on the world-sheet of the probe string can be identified with the radial variable of the ten-dimensional metric, see \eqref{eq:10dmetric}, i.e. $\sigma^2=r$. The coupling to the scalar field in $\mathcal{N}=2^*$ in \eqref{eq:WLdef} is manifested in the bulk by the probe string having a profile along the $S^5$ directions in the ten-dimensional metric. Since translations along $t$ are a symmetry of the ten-dimensional solution we make the reasonable assumption that the induced fields only depend on $r$ and not on $t$. This means that since $B^{(2)}$ has legs only along the internal $S^5$ directions we immediately find that $P[B^{(2)}]=0$. The induced metric takes the form
 \begin{align}
 P[\dd s^2_{10}] = \frac{(CK_1K_2)^{1/4}}{\eta\sqrt{g_s}}\bigg[\left(1+G_{mn}\frac{\dd \Theta^m}{\dd r}\frac{\dd \Theta^n}{\dd r}\right)\dd r^2 + \e^{2A}\dd t^2  \bigg]\,,
 \end{align}
 where $G_{mn}$ is the metric on the deformed $S^5$ given in \eqref{eq:S5deformed} and the functions $\Theta^m(r)$ describe the profile of the string world-sheet on $S^5$. We can identify the functions $\Theta^m$ with the five coordinates $\{\theta,\phi,\omega,\xi_{+},\xi_{-}\}$ in \eqref{Eq: Round five sphere metric}. Minimizing the string action in \eqref{eq:Sstrdef} amounts to minimizing
\begin{equation}\label{eq:detPG}
\e^\Phi \det P[G_{MN}] =  \eta^{-2}\e^{2A}\left( CK_1 \sin^2\phi + K_2 \cos^2\phi\right)\left(1+G_{mn}\dot \Theta^m \dot \Theta^n\right)\,,
\end{equation}
where we use a dot to denote the $r$-derivative of $\Theta^m$. Notice that $G_{mn}\dot \Theta^m \dot \Theta^n$ is a sum of non-negative terms and thus the expression in \eqref{eq:detPG} can be minimized by setting each term to zero. This is achieved by keeping the position of the string on the $ S^5$ independent of $r$, i.e. $\dot \Theta^m=0$. The position of the string on $ S^5$ is then obtained by minimizing the resulting function
\begin{equation}\label{eq:detPGconst}
\left.\e^\Phi \det P[G_{MN}]\right|_{\dot \Theta^m=0} =  \eta^{-2}\e^{2A}\left( CK_1 \sin^2\phi + K_2 \cos^2\phi\right)\,,
\end{equation}
along the coordinates of $ S^5$. It is not hard to see that the minima of \eqref{eq:detPGconst} are at:\footnote{These minima can also be found by using the Euler-Lagrange equations derived from \eqref{eq:Sstrdef} and \eqref{eq:detPG}.}
\begin{equation}\label{eq:minima}
\theta = \frac{\pi}{2}\,,\qquad \phi= \frac{n\pi}{2}~,\qquad n=0,1,2,3\,.
\end{equation}
Only one of these minima, $\phi=0$, corresponds to the Wilson line operator in \eqref{eq:WLdef}. This can be shown by looking at the corresponding position in the embedding space of the five sphere, see \eqref{Eq: Embedding coordinates of five sphere}
\begin{equation}
Y^1=Y^2=Y^3=Y^4=0~,\quad Y^5 = \cos\frac{n\pi}{2}~,\quad Y^6 = -\sin\frac{n\pi}{2}~,\qquad n=0,1,2,3~.
\end{equation}
This implies that the scalar coupling in the dual Wilson loop operator \eqref{eq:WLdef} involve only $X_3$ or $X_6$ for $n=0,2$ and a linear combination of $X_3$ and $X_6$ for $n=1,3$. Since the localization setup is adapted to calculate the vev of the Wilson loop operator defined in \eqref{eq:WLdef} with a coupling to $X_3$ we focus exclusively on the probe string sitting at the $n=0$ minimum. 

The string action for $n=0$ reduces to 
\begin{align}\label{Eq: Reduced string world sheet action}
S_\text{string} =   \frac{1}{\ell_s^2} \int \dd r \; \e^{A} \eta^2 =  \sqrt{\lambda} \int \dd \tilde r \; \e^{\tilde A} \eta^2\,,
\end{align}
where we have performed the integral over the great circle on $S^4$ parametrized by $t$, used the dimensionless variables $\tilde r = r/L_\text{AdS}$ and $\e^{\tilde A} = \e^{A}/L_\text{AdS}$, and used the relation in \eqref{eq:constants}.

We now proceed to evaluate the on-shell action of the probe string described above. Using the UV expansion in \eqref{eq:UV5d} one finds that the integrand in \eqref{Eq: Reduced string world sheet action} diverges close to the boundary as
\begin{align}
\lim\limits_{\tilde r\rightarrow \infty} \eta^2 \e^{\tilde A} = \frac{1}{2} \e^{\tilde r} + {\cal O}(1)\,.
\end{align}
We thus have to regularize the action in \eqref{Eq: Reduced string world sheet action} by adding suitable counterterms. The counterterms should be covariant quantities on the world-sheet boundary, given by the contour $\mathcal{C}$, built from the ten-dimensional background fields, see for example \cite{Karch:2005ms,Chu:2008xg}. The only such non-vanishing quantity is a ``cosmological constant'' on the world-sheet boundary. Notice that the determinant of the (dimensionless) metric on the world-sheet boundary is just $\eta~ \e^{\tilde A}$ and using \eqref{eq:UV5d} we find that the counterterm in question has the asymptotic behavior
\begin{align}\label{counterterm}
\lim\limits_{r\rightarrow\infty} \eta~ \e^{\tilde A} =  \frac{1}{2}\e^{\tilde r} + {\cal O}(\tilde r\e^{-\tilde r})\,.
\end{align}
Other possible covariant counter terms are given by a positive power, $k$, of the ten-dimensional dilaton evaluated on the boundary. One can use \eqref{eq:10dPhiC0} and \eqref{eq:UV5d} to show that these terms approach zero as $e^{-\tilde r}$ or faster:
\begin{align}\label{Eq: Additional possible counter terms}
\lim\limits_{\tilde r\rightarrow\infty }\eta~ \e^{\tilde A} \,\Phi^k=&  \, \mathcal{O}(\e^{-(2k-1)\tilde r})\,.
\end{align}
We therefore conclude that the only infinite or finite counterterm that can be used to regularize the string action in \eqref{Eq: Reduced string world sheet action} is the one in \eqref{counterterm}. Note that this counterterm gives identical results for the regularized on-shell action as the background subtraction method which is often employed in holographic Wilson line calculation\cite{Drukker:1999zq}. 

\begin{figure}
\centering
\begin{subfigure}{.5\textwidth}
  \centering
  \includegraphics[width=.8\linewidth]{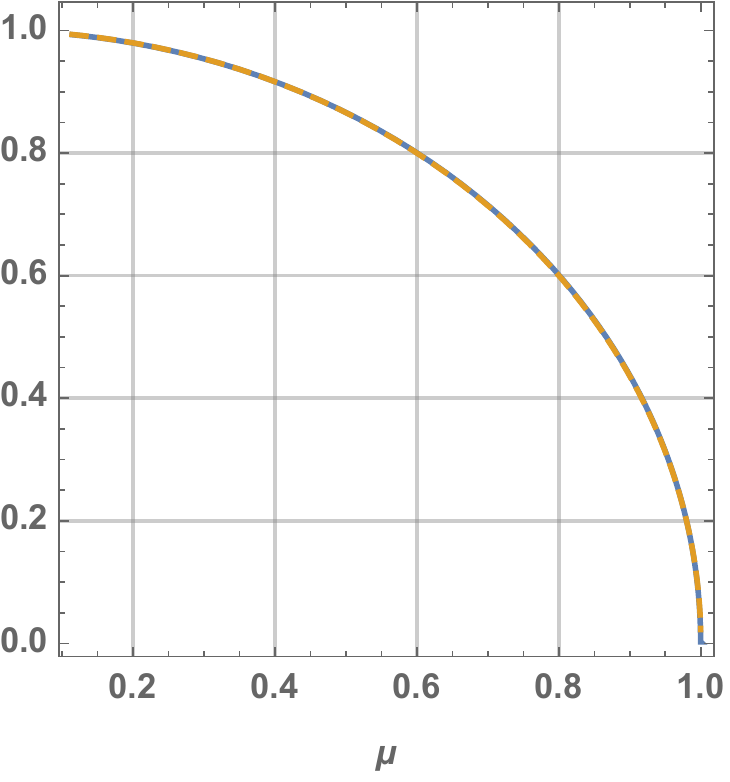}
  \caption{$\mu$ real.}
  \label{fig:sub1}
\end{subfigure}%
\begin{subfigure}{.5\textwidth}
  \centering
  \includegraphics[width=.8\linewidth]{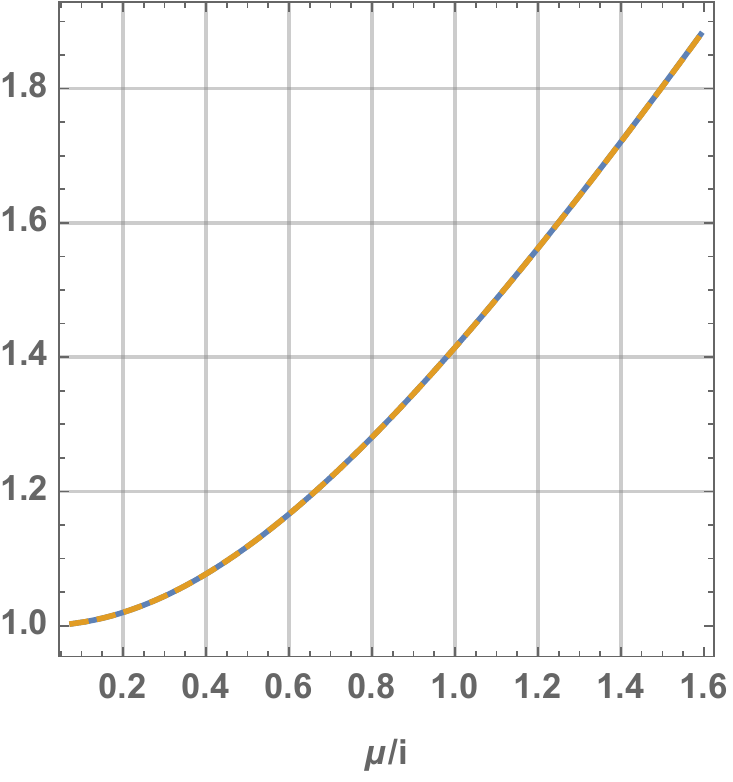}
  \caption{$\mu$ imaginary.}
  \label{fig:sub2}
\end{subfigure}
\caption{The yellow dashed curve shows numerical solutions for the on-shell action in \eqref{Eq: Renormalized world sheet action 1} divided by $\sqrt{\lambda}$ as a function of the parameter $\mu$. The blue solid curve is a plot of the function $\sqrt{1 - \mu^2}$ which should be compared to the field theory result in \eqref{eq:Wvev}.\label{Fig:WLlocalisation}}
\label{fig:test}
\end{figure}

The renormalized on-shell action of the probe string of interest is thus given by
\begin{align}\label{Eq: Renormalized world sheet action 1}
S_\text{string}^R = \lim_{\tilde r_\text{UV}\to \infty}\sqrt{\lambda}\left(\int_{\tilde r_*}^{\tilde r_\text{UV}} \dd \tilde r\, e^{\tilde A} \eta^2 - e^{\tilde A(\tilde r_\text{UV})} \eta(\tilde r_\text{UV}) \right)\,,
\end{align}
and, as expected, it is independent of the UV cutoff scale $\tilde r_\text{UV}$. Notice that the integration range in \eqref{Eq: Renormalized world sheet action 1} extends all the way to the IR cutoff $r_*$ in the bulk where the five-dimensional metric caps off. This is because the string sits on the equator of $S^4$ for all values of the radial coordinate $r$ in the bulk geometry. We are now ready to use the renormalized action in \eqref{Eq: Renormalized world sheet action 1} to determine the vev of the dual Wilson loop, employing the relation in \eqref{Eq: General Wilson loop dictionary}.
Since the ten-dimensional solution described in Section~\ref{Sec: A new solution of type IIB supergravity} can only be obtained numerically we cannot aim at reproducing the localization result in \eqref{eq:Wvev} analytically. We can however use the numerical approach outlined above \eqref{eq:vmurelation} and construct many numerical solutions for different values of the parameter $\mu$ in \eqref{eq:UV5d}. For each of these numerical supergravity solutions we can then evaluate the regularized probe string action in \eqref{Eq: Renormalized world sheet action 1}. As discussed above \eqref{eq:vmurelation} we have focused on real or pure imaginary values of $\mu$. Our numerical results for the regularized string action are plotted in Figure~\ref{fig:test}. It is clear from this figure that, upon using the identification between the supergravity parameter $\mu$ and the dimensionless mass, $ma$, see \eqref{eq:muimadef}, we find an excellent agreement between our numerical supergravity results and the supersymmetric localization result in \eqref{eq:Wvev}. Finally we want to stress that we have checked that our numerical results for the regularized on-shell action in \eqref{Eq: Renormalized world sheet action 1} are in very good agreement with the function $\sqrt{1-\mu^2}$ also for general values of the complex parameter $\mu$ away from the real and imaginary axis.

\section{Conclusions}
\label{sec:conclusions}

In this paper we studied the ten-dimensional supergravity dual of the $\mathcal{N}=2^*$ theory on $S^4$. To this end we used the five-dimensional supergravity dual of this gauge theory constructed in \cite{Bobev:2013cja} and the explicit uplift formulae of \cite{Baguet:2015sma} to construct a new supersymmetric solution of type IIB supergravity. We then explored this new background with probe fundamental strings which are holographically dual to Wilson lines in the fundamental representation of the gauge group. We identified a simple circular probe string which is dual to the supersymmetric Wilson line in \eqref{eq:WLdef}. The vev of this Wilson line can be calculated analytically in the planar limit at large 't Hooft coupling by supersymmetric localization. We showed that the regularized on-shell action of the probe string we consider is equal to the vev of this supersymmetric Wilson line for general values of the mass parameter in the $\mathcal{N}=2^*$ theory. This constitutes a non-trivial test of holography for non-conformal gauge theories. Our work naturally leads to several interesting questions which can be explored in the near future.

We have not rigorously shown that the probe string we studied in Section~\ref{Sec: Wilson loops and strings} is indeed supersymmetric. It should be possible to do so using $\kappa$-symmetry.\footnote{A recent $\kappa$-symmetry analysis for probe branes in asymptotically AdS$_5$ supergravity solutions with curved boundaries was performed  in \cite{Karch:2015vra,Robinson:2017sup} for probe D5- and D7-branes.} In addition it will be interesting to explore other supersymmetric Wilson lines in the fundamental representation in the $\mathcal{N}=2^{*}$ theory on $S^4$. These operators are different from the one in \eqref{eq:WLdef} since they involve also the scalar $X_6$. To the best of our knowledge their vevs have not been calculated via supersymmetric localization. Nevertheless our new type IIB supergravity background should allow to compute these vevs, at least for large $\lambda$, by studying suitable probe fundamental strings. 

We have limited our holographic analysis to Wilson loop operators in the fundamental representation. There are other supersymmetric Wilson loops in the $\mathcal{N}=2^{*}$ theory on $S^4$ whose vevs can be computed in the planar limit using supersymmetric localization, see for example \cite{Chen-Lin:2015dfa,Chen-Lin:2015xlh,Liu:2017fiq,Russo:2017ngf}. For large values of $\lambda$ these field theory calculations should be compared with supergravity. This can be done by studying appropriate probe D3- and D5-branes in the type IIB supergravity solution discussed in Section~\ref{Sec: From five to ten dimensions}. The vev of the corresponding Wilson loop operators can then be computed by evaluating the regularized on-shell action of these probe branes. Much less is known about supersymmetric 't Hooft line operators or surface operators in the $\mathcal{N}=2^*$ theory. From the supergravity perspective these operators should correspond to probe D1- and D3-branes, respectively and it should be possible to classify them systematically and compute their expectation values holographically. It will certainly be interesting to explore this further.

 The planar limit of the $\mathcal{N}=2^*$ SYM theory on $S^4$ exhibits a rich structure of phase transitions as one varies the 't Hooft coupling $\lambda$. As we demonstrated in this paper the supergravity solution discussed in Section~\ref{Sec: From five to ten dimensions} is clearly well-suited for studying holographically the gauge theory for $\lambda \gg 1$. It is desirable however to develop techniques from string theory and supergravity which allow us to compute $1/\sqrt{\lambda}$ corrections to physical observables in the planar $\mathcal{N}=2^*$ SYM theory. Some results in this direction were derived in  \cite{Chen:2014vka,Chen-Lin:2017pay} but it is certainly very interesting to study this further since it offers the exciting possibility to connect exact localization calculations for non-conformal gauge theories with string theory corrections to supergravity.
 
The uplift formulae in  \cite{Baguet:2015sma} provide a concrete tool to construct explicit solutions of type IIB supergravity by first solving the equations of motion of the five-dimensional $\mathcal{N}=8$ $\SO (6)$ gauged supergravity. Obtaining such five-dimensional solutions is usually a much simpler enterprise then directly solving the equations of motion of type IIB supergravity. It is thus natural to apply these powerful technical results to construct explicit uplifts of other interesting five-dimensional solutions which have found applications in holography. Two particular examples which should be easily accessible are the gravity dual of the $\mathcal{N}=1^*$ cousin of the $\mathcal{N}=2^*$ theory on $\mathbb{R}^4$ \cite{Girardello:1999bd,Pilch:2000fu} and $S^4$ \cite{Bobev:2016nua}.

 \newpage
\bigskip
\bigskip
\leftline{\bf Acknowledgements}
\smallskip
\noindent We are grateful to Yago Bea, Ioannis Papadimitriou, and in particular to Diego Trancanelli for initial collaboration on some of the results presented here and interesting discussions. In addition we would like to thank Henriette Elvang, Ben Niehoff, Silviu Pufu, and Kostya Zarembo for useful conversations. The work of NB is supported in part by an Odysseus grant G0F9516N from the FWO. FFG is a Postdoctoral Fellow of the Research Foundation - Flanders. JvM is a PhD Fellow of the Research Foundation - Flanders. We are also supported by the KU Lueven C1 grant ZKD1118 C16/16/005, and by the Belgian Federal Science Policy Office through the Inter-University Attraction Pole P7/37.  
 
 \appendix
 \section{Spherical coordinates}
 \label{Sec: Coordinates}

Here we collect some details on the coordinates on $S^4$ and $S^5$ that we used throughout the paper.

For the five-dimensional metric in \eqref{eq:5dmetric} we employ the following Einstein metric on $S^4$ 
\begin{align}\label{eq:S4met}
 \dd \Omega^2_{4}= \dd\zeta_1^2 +\sin ^2\zeta_1 \left(\dd\zeta_2^2+\sin^2\zeta_2 \left(\dd\zeta_3^2+\sin^2\zeta_3\dd\zeta_4^2\right)\right)\,.
 \end{align}

Throughout our calculations we have used the following explicit embedding of $S^5$ in $\mathbb{R}^6$: 
\begin{equation}\label{Eq: Embedding coordinates of five sphere}
  \begin{split}
    Y^1 &=\cos\theta\cos\frac{1}{2}(\xi_{+}+\xi_{-})\cos\frac{\omega}{2}\,,\\
    Y^3 &= -\cos\theta\sin\frac{1}{2}(\xi_{+}+\xi_{-})\cos\frac{\omega}{2}\,,\\
    Y^5 &= \sin\theta\cos\phi\,,
  \end{split}
\qquad
  \begin{split}
    Y^2 &=\cos\theta\sin\frac{1}{2}(\xi_{+}-\xi_{-})\sin\frac{\omega}{2}\,,\\
    Y^4 &=\cos\theta\cos\frac{1}{2}(\xi_{+}-\xi_{-})\sin\frac{\omega}{2}\,,\\
    Y^6 &=-\sin\theta\sin \phi\,.
  \end{split}
\end{equation}
The choice of coordinates in \eqref{Eq: Embedding coordinates of five sphere} is adapted to the symmetries of the problem. Namely the coordinates $\{Y^{1},Y^{2},Y^{3},Y^{4}\}$ can be formally identified with the two-dimensional complex plane spanned by the scalars $Z_{1,2}$ in the adjoint hypermultiplet of $\mathcal{N}=4$ SYM, see \eqref{eq:hypmultdef}. Similarly the coordinates $\{Y^{5},Y^{6}\}$ should be identified with the complex plane spanned by the scalar $Z_3$ in \eqref{eq:vecmultdef}. These identifications imply that the $\U(1)_V\times \U (1)_H\times \U(1)_R$ symmetries discussed below \eqref{eq:LmasS4} should be identified with the Killing vectors $\partial_{\xi_+}$, $\partial_{\xi_-}$, and $\partial_{\phi}$, respectively.
In these coordinates, the Einstein metric on $S^5$ reads
\begin{align}\label{Eq: Round five sphere metric}
\dd\widehat{\Omega}_{5}^2=\dd\theta^2+\sin^2\theta \,\dd\phi^2+\cos^2\theta \left(\sigma_1^2+\sigma_2^2+\sigma_3^2\right)\,,
\end{align}
where $\sigma_i$ are the $\SU(2)$ left-invariant one-forms which obey the relation:
\begin{align}
\dd \sigma_i =  \varepsilon_{ijk} \sigma_j \wedge \sigma_k\,.
\end{align}
Explicitly the one-forms are given by
 \begin{align}\label{eq:sig_i}
 \sigma_1 =& - \frac{1}{2} \left(\sin\xi_{-}\dd\omega-\sin \omega  \cos \xi_{-} \dd\xi_{+} \right)\,,   \notag\\
 \sigma_2 =& -\frac{1}{2} \left(\cos\xi_{-}\dd\omega+\sin  \omega  \sin \xi_{-}\dd\xi_{+}  \right)\,, \\	
 \sigma_3 =& -\frac{1}{2} \left(\dd\xi_{-}+\cos \omega \dd\xi_{+}\right)\,. \notag
 \end{align}
To avoid conical singularities the range of the coordinates on $S^5$ should be
\begin{equation}
\theta\in[0,\pi/2]\,, \qquad \omega\in[0,\pi]\,, \qquad  \xi_{+},\xi_{-},\phi\in[0,2\pi]\,.
\end{equation}

To apply the uplift formulae of \cite{Baguet:2015sma} discussed in Section~\ref{Sec: From five to ten dimensions}  we also need the 15 Killing vectors of the round five-sphere which can be obtained directly from the six embedding functions $Y_I$ via
\begin{align}\label{Eq: Killing vectors of round five sphere}
\mathcal{K}^{\phantom{IJ}m}_{IJ} = -g \,\widehat G^{mn} Y_{[I} \nabla_{n} Y_{J]}\,.
\end{align}
The derivative in \eqref{Eq: Killing vectors of round five sphere} is with respect to the coordinates on the five-sphere, $\{\theta, \phi, \omega, \xi_{+}, \xi_{-}\}$, and $\widehat G^{mn}$ is the inverse of the metric in \eqref{Eq: Round five sphere metric}.

 \section{5d $\mathcal{N}=8$ supergravity}
 \label{Sec: Some properties of the exceptional group E66}
 
Here we collect some formulae from \cite{Gunaydin:1985cu} that pertain to our discussion. The scalar potential of the five-dimensional supergravity action in \eqref{Eq: Baguet:2015sma Lagrangian} is given by
\begin{equation}\label{eq:5dpotential}
V = - \frac{g^2}{32}\left(2W_{ab}W^{ab}-W_{abcd}W^{abcd}\right)\,,
\end{equation}
where the $W$ tensors are defined as
\begin{equation}
W_{abcd} = \epsilon^{\alpha\beta}\delta^{IJ}V_{I\alpha ab}V_{J\beta cd}\,, \qquad W_{ab} = W^{c}\,_{acb}\,.
\end{equation}
The $V_{I\alpha ab}$ are related to the group element $U$ introduced below \eqref{Eq: Xhat matrix} through a change of basis. Specifically, the $V_{I\alpha ab}$ has curved indices $I\alpha$ in an $\SL(6,\mathbb{R})\times \SL(2,\mathbb{R})$ representation, while the flat indices $ab$ are in an $\USp(8)$ representation. The $U$ matrices on the other hand are written entirely in the $\SL(6,\mathbb{R})\times \SL(2,\mathbb{R})$ basis. The relation between the two objects is given by a set of gamma matrices:
\begin{align}
\Gamma_{I\alpha} = \left(\Gamma_{I}, \rmi \Gamma_{I} \Gamma_{0} \right)\,,\quad \text{and} \quad \Gamma_{IJ} = \Gamma_{[I} \Gamma_{J]}\,,
\end{align}
which transform under $\SO(7)$ and have the following properties
\begin{align}
\left\{\Gamma_{I},\Gamma_{J} \right\} = 2\delta_{IJ}\,,\quad \text{and} \quad \Gamma_{0}\ldots \Gamma_{6} = - \rmi \mathbb{1}\,.
\end{align} 
The explicit relation between the frame fields is then given by the following expression:
\begin{equation}
\begin{aligned}
V_{I\alpha}^{\quad ab} =& \frac14 \frac{1}{\sqrt{2}} \left( \Gamma_{KL}^{\quad\, ab} U_{I \alpha\, KL} + 2 \Gamma_{K\beta}^{\quad\: ab} U_{I\alpha}^{\quad K\beta} \right).
\end{aligned}
\end{equation}
More information about the relation between $\USp(8)$ and $\SL(6,\mathbb{R})\times \SL(2,\mathbb{R})$ representations and their embedding in $\E_{6(6)}$ can be found in \cite{Gunaydin:1985cu}.

\section{Type IIB supergravity}
\label{app:IIBsugra}

To test the uplift formulae of \cite{Baguet:2015sma}, given in Section~\ref{Sec: From five to ten dimensions}, we have checked explicitly that all ten-dimensional supergravity backgrounds discussed in this paper satisfy the equations of motion of type IIB supergravity. In this appendix we give the definition of the field strengths and the equations of motion of the ten-dimensional theory.

The NS-NS three-form flux is defined as 
\begin{align}
H^{(3)} = \dd B^{(2)}\,.
\end{align}
The R-R field strengths are defined as 
\begin{equation}\label{Eq: RR fields on internal space}
\begin{aligned}
F^{(1)} &= \dd C_0\,, \qquad F^{(3)} = \dd C^{(2)} - C_0 H^{(3)}\,,\\
\mathcal{F}^{(5)} &= \dd C^{(4)} - \frac12\left( C^{(2)} \wedge H^{(3)} - B^{(2)} \wedge \dd C^{(2)} \right)\,.
\end{aligned}
\end{equation}
Notice that $C^{(4)}$ calculated via the uplift formula \eqref{eq:C4uplift} only provides \emph{half} of the type IIB five-form. The full five-form flux must be self-dual and is thus given by
\begin{align}
F^{(5)} = \mathcal{F}^{(5)} + \star_{10} \mathcal{F}^{(5)}\,,
\end{align}
where $\star_{10}$ is the ten-dimensional Hodge dual. 

The equations of motion are given by the set of Maxwell equations and Bianchi identities for the NS-NS sector
 \begin{equation}
 \begin{aligned}
&\dd H^{(3)} = 0\,,\quad
&\dd\left( \star_{10}e^{-\Phi} H^{(3)}\right) - F^{(3)} \wedge {F}^{(5)} - \rme^{\Phi}F^{(1)} \wedge\star_{10}F^{(3)} = 0\,,
\end{aligned}
\end{equation}
and similarly for the R-R sector 
\begin{equation}
\begin{aligned}
&\dd F^{(1)} = 0\,,\quad
& \dd \left( \star_{10} \rme^{2\Phi} F^{(1)} \right) +  \rme^{\Phi} H^{(3)} \wedge \star_{10} F^{(3)} =0\,,  \\
&\dd F^{(3)} - H^{(3)} \wedge F^{(1)} = 0\,, \quad
&\dd \left(\star_{10} \rme^{\Phi}F^{(3)}\right) + H^{(3)} \wedge {F}^{(5)} = 0\,, \\
& \dd  F^{(5)}  -  H^{(3)} \wedge F^{(3)}= 0\,,\quad 
&
\end{aligned}
\end{equation}
the dilaton equation
\begin{align}
 \Delta \Phi +  \frac12\rme^{-\Phi} |H^{(3)}|^2 -  \rme^{2\Phi} |F^{(1)}|^2 - \frac12 \rme^{\Phi} |F^{(3)}|^2 =0\,,
 \end{align}
and the ten-dimensional Einstein equation
\begin{equation}
 \begin{aligned}
 &R_{MN} - \frac12 \nabla_{M}\Phi \nabla_{N} \Phi -\frac12 \rme^{-\Phi}  |H^{(3)}|^2_{MN} - \frac12 \e^{2\Phi}  |F^{(1)}|^2_{MN}  - \frac12 \e^{\Phi}  |F^{(3)}|^2_{MN} \\
 &  - \frac14   |F^{(5)}|^2_{MN} + \frac18 g_{MN} \left(\rme^{-\Phi}  |H^{(3)}|^2+\e^{\Phi}|F^{(3)}|^2    +   |F^{(5)}|^2 \right) =0  \,.
 \end{aligned}
\end{equation}
We have introduced the notation
\begin{equation}
|F^{(p)}|^2 = \frac{1}{p!}F^{(p)}_{M_1M_2\cdots M_p}F^{(p)M_1M_2\cdots M_p}~,\qquad |F^{(p)}|^2_{MN} = \frac{1}{(p-1)!}F^{(p)}_{M~M_1\cdots M_{p-1}}F^{(p)M_1\cdots M_{p-1}}_N~.
\end{equation}
Note that the equations above are presented in Einstein frame. The relation between Einstein and string frame is given by a rescaling of the metric, $G^{({\rm string})}_{MN} = \rme^{\Phi/2}G^{({\rm Einstein})}_{MN}$.

\section{Pilch-Warner and Coulomb branch solutions}
\label{app:PWSTU}

To gain some confidence that we have applied the uplift formulae of \cite{Baguet:2015sma} correctly it is instructive to reproduce two well-known solutions of type IIB supergravity which can be obtained as limits of the three-scalar model in Section~\ref{subsec:3scalar}. The limit we want to take is to first replace the $S^4$ in \eqref{eq:5dmetric} by $\mathbb{R}^4$. Then one can show that it is consistent to set $\beta=0$, or equivalently $T=0$, in the BPS equations in \eqref{Eq: BPS equations for C and T}. In this limit one can integrate the BPS equations analytically.

\subsection{Pilch-Warner}

The type IIB supergravity solution found in \cite{Pilch:2000ue}  is the holographic dual of the $\mathcal{N}=2^*$ theory on $\mathbb{R}^4$. It can be obtained by taking $T=0$ and keeping the other two scalars in Section~\ref{subsec:3scalar} non-trivial. This procedure is equivalent to taking the limit $z=-\tilde{z}$. The functions in \eqref{Eq: Scalar functions} then become\footnote{Note that $K_1$ and $K_2$ correspond to the functions $X_1$ and $X_2$ in \cite{Pilch:2000ue}. The scalar $C$ is simply equal to $\cosh 2\chi$, which is in harmony with the conventions in \cite{Pilch:2000ue}.}
\begin{equation}
\begin{aligned}\label{Eq: Scalar functions in PW}
S =& 1,\quad
K_1 = \cos^2\theta + \eta^6 C\sin^2\theta ,\quad
K_2 = C \cos^2\theta + \eta^6\sin^2\theta .
\end{aligned}
\end{equation}
The full type IIB supergravity background obtained using the uplift formulae of \cite{Baguet:2015sma} is then given by 
\begin{equation}
\begin{aligned}
\dd \Omega_{5}^2 =& \frac{4}{g^2 \eta^2} \Bigg(\frac{\dd\theta^2}{C}+\eta^6\cos^2\theta  \left(\frac{\sigma_1^2+\sigma_2^2}{K_1}+\frac{\sigma_3^2}{K_2C}\right)+\frac{\sin^2\theta \dd\phi^2}{K_2}\Bigg),\\
e^{\Phi} =& \frac{g_s}{\sqrt{C K_1 K_2}} (C K_1 \sin^2 \phi + K_2 \cos^2 \phi), \qquad  C^{(0)}= \frac{\eta^6\left(C^2 -1\right)\sin^2\theta\sin2\phi }{2 g_s \left(C K_1 \sin^2\phi+K_2 \cos^2\phi\right)},\\
B^{(2)}= &-\frac{2\sqrt{C^2 -1}}{g^2 }\sin2\theta \cos\theta \cos\phi\Bigg( \frac{2}{ C\sin 2\theta }  \dd\theta\wedge \sigma_3 + \frac{\tan\phi}{K_2} \sigma_3 \wedge \dd\phi + \frac{\eta^6}{K_1} \sigma_2 \wedge \sigma_1 \Bigg),\\
C^{(2)}= &\frac{2\sqrt{C^2-1} }{g_s g^2}\sin2\theta\cos\theta\sin\phi \Bigg( \frac{2}{C \sin 2\theta }  \dd\theta\wedge \sigma_3 - \frac{ \cot\phi}{K_2} \sigma_3 \wedge \dd\phi + \frac{\eta^6}{K_1} \sigma_2 \wedge \sigma_1 \Bigg),\\
C^{(4)} =&\frac{ 8\left( C K_1 + K_2 \right)}{g_s g^4  K_1 K_2}  \cos^4\theta \,  \sigma_1 \wedge \sigma_2 \wedge \sigma_3 \wedge \dd \phi. 
\end{aligned}
\end{equation}
Notice that in this limit we Wick rotate to Lorentzian signature and the five-dimensional metric is a flat Minkowski domain wall:
\begin{equation}\label{eq:flatslice}
\dd s_{1,4}^2 = \dd r^2 + e^{2A} \eta_{\mu\nu} \dd x^{\mu}\dd x^{\nu}\,,
\end{equation}
where $A$ does not obey the algebraic BPS equation given in \eqref{eq:e2Aalg} since when $z=-\tilde z$ the algebraic expression is not valid. One can obtain an expression for $A$ from the differential equation it satisfies \eqref{eq:Adiff}. The explicit form for $A$ is
\begin{equation}
\begin{aligned}
\e^{2A} = \frac{\eta^4}{C^2-1}\,.
\end{aligned}
\end{equation}
The full ten-dimensional type IIB metric is then
\begin{equation}
\begin{aligned}
\dd s_{10}^2 = \frac{\left(C K_1 K_2\right)^{1/4}}{\eta \sqrt{g_s}} \left( \dd s_{1,4}^2 + \dd \Omega_{5}^2 \right)\,.
\end{aligned}
\end{equation}
We have verified that this background matches the one derived in \cite{Pilch:2000ue} upon setting $g_s=1$ and changing to mostly minus signature. Note that our $\eta$ is their $\rho$, our $\sigma_1$ is their $\sigma_3$ and vice versa,  and their complex two-form $A^{(2)}$ is given by
\begin{align}
A^{(2)} = C^{(2)} + \rmi B^{(2)}\,,
\end{align}
in our notation. There are two minor typos in \cite{Pilch:2000ue}. In Equation (3.28) of \cite{Pilch:2000ue} $a_3$ should be $-a_3$ and in Equation (4.9) the denominator should be $4 \sinh^4(2\chi)$.

\subsection{Coulomb branch}

An even simpler solution of type IIB supergravity can be obtained by setting $\beta,\chi=0$, or alternatively $C=1,~ T=0$. This background was studied in \cite{Cvetic:1999xx} and is holographically dual to an RG flow on the vacuum moduli space of $\mathcal{N}=4$ SYM, sometimes referred to as Coulomb branch flow, where a single operator sitting in the $\mathbf{20'}$ representation of $\SO (6)$ acquires a vev. To compare our conventions to the ones in \cite{Cvetic:1999xx} we should relate the scalars of the five-dimensional supergravity in Equation (2.4) of \cite{Cvetic:1999xx} to ours in the following way:
\begin{align}
X_1 = X_2 = X_3 = X_4 = \frac{1}{\eta^2}\,, \quad \text{and} \quad X_5 = X_6 = \eta^4\,.
\end{align}
For this set of scalars the functions in \eqref{Eq: Scalar functions} simplify to
\begin{align}
S=1, \qquad K_1 = K_2 = K = \cos^2\theta + \eta^6 \sin^2\theta\,.
\end{align}
Using the uplift formulae in \cite{Baguet:2015sma} one can then find the following simple solution of type IIB supergravity 
\begin{equation}
\begin{aligned}
\dd \Omega_5^2=& ~\frac{4}{g^2 \eta^2}\bigg(  \dd\theta^2+ \frac{\eta^6\cos^2\theta }{K} \left(\sigma_1^2+\sigma_2^2+\sigma_3^2\right)+\frac{ \sin ^2\theta}{K}\dd\phi^2\bigg)\,,\\
C^{(4)} =&~\frac{16 \cos^4\theta}{g_s g^4 K}   \sigma_1 \wedge \sigma_2 \wedge \sigma_3\wedge\dd \phi, \\
\e^\Phi=&g_s~,\qquad C^{(0)}=B^{(2)}=C^{(2)}= 0\,.
\end{aligned}
\end{equation}
Once again the domain wall is sliced by four-dimensional Minkowski space and the five-dimensional metric is given by \eqref{eq:flatslice} where $A$ and the scalar $\eta$ are determined by
\begin{equation}
\begin{aligned}
A' =\frac{g}{6}(\eta^4+2\eta^{-2})\,,\qquad (\log\eta)' = -\frac{g}{6}\left(\eta^4-\eta^{-2}\right)\,.
\end{aligned}
\end{equation}
One can then solve for the metric function $A$ in terms of the scalar $\eta$ and find
\begin{equation}
\begin{aligned}
\e^{2A}=\eta^2-\eta^{-4}\,.
\end{aligned}
\end{equation}
Finally the full ten-dimensional metric is given by
\begin{equation}
\begin{aligned}
\dd s_{10}^2 = \frac{\sqrt{K}}{\eta \sqrt{g_s}} \left( \dd s_{1,4}^2 + \dd \Omega_{5}^2 \right)\,.
\end{aligned}
\end{equation}

\bibliography{N2starIIB}
\bibliographystyle{JHEP}

\end{document}